\documentclass[aps,a4paper,twocolumn,10pt, accepted=2023-05-11]{quantumarticle}
\pdfoutput=1

\usepackage[utf8]{inputenc}
\usepackage[english]{babel}
\usepackage[T1]{fontenc}
\usepackage[normalem]{ulem}

\usepackage[SquareTraceBrackets]{quantum}
\usepackage[numbers,sort&compress]{natbib}

\usepackage{graphicx,mathrsfs,amsmath,amsbsy,bm,optidef}
\usepackage{optidef}

\usepackage[dvipsnames]{xcolor}
\definecolor{myblue}{named}{MidnightBlue}

\usepackage{amsthm}

\usepackage{xr-hyper}

\usepackage{multirow}
\usepackage{enumitem}
\usepackage{listings}
\usepackage[caption=false]{subfig}
\usepackage{tabularx}
\DeclareGraphicsExtensions{.pdf, .jpg, .eps, .svg, .png}
 
\usepackage[colorlinks=true,citecolor=myblue,linkcolor=myblue,urlcolor=myblue]{hyperref}
\usepackage{cleveref}

\makeatletter
\def\thickhline{%
  \noalign{\ifnum0=`}\fi\hrule \@height \thickarrayrulewidth \futurelet
   \reserved@a\@xthickhline}
\def\@xthickhline{\ifx\reserved@a\thickhline
               \vskip\doublerulesep
               \vskip-\thickarrayrulewidth
             \fi
      \ifnum0=`{\fi}}
\makeatother

\begin{document}

\title{Linear optics and photodetection achieve near-optimal unambiguous coherent state discrimination}

\author{Jasminder S. Sidhu}
\email{jsmdrsidhu@gmail.com}
\affiliation{SUPA Department of Physics, The University of Strathclyde, Glasgow, G4 0NG, UK}
\orcid{0000-0002-6167-8224}

\author{Michael S. Bullock}
\affiliation{Department of Electrical and Computer Engineering, The University of Arizona, Tucson, Arizona 85721, USA}
\orcid{0000-0002-3528-7473}

\author{Saikat Guha}
\affiliation{Department of Electrical and Computer Engineering, The University of Arizona, Tucson, Arizona 85721, USA}
\affiliation{College of Optical Sciences, The University of Arizona, Tucson, Arizona 85721, USA}
\orcid{0000-0002-2581-4380}

\author{Cosmo Lupo}
\email{cosmo.lupo@poliba.it}
\affiliation{Dipartimento Interateneo di Fisica, Politecnico \& Universit\`a di Bari, 70126 Bari, Italy}
\affiliation{INFN, Sezione di Bari, 70126 Bari, Italy}
\orcid{0000-0002-5227-4009}

\begin{abstract}
\noindent Coherent states of the quantum electromagnetic field, the quantum description of ideal laser light, are prime candidates as information carriers for optical communications. A large body of literature exists on their quantum-limited estimation and discrimination. However, very little is known about the practical realizations of receivers for unambiguous state discrimination (USD) of coherent states. Here we fill this gap and outline a theory of USD with receivers that are allowed to employ: passive multimode linear optics, phase-space displacements, auxiliary vacuum modes, and on-off photon detection. Our results indicate that, in some regimes, these currently-available optical components are typically sufficient to achieve near-optimal unambiguous discrimination of multiple, multimode coherent states.
\end{abstract}

\maketitle


\section{Introduction}
\label{sec:intro}

\noindent
Quantum mechanics places fundamental limits on the distinguishability of non-orthogonal quantum states. This fact underpins applications in quantum information science and technology, most notably in quantum cryptography~\cite{Bennett1992_PRL}, and constrains the performance of quantum sensors~\cite{Sidhu2020_AVS, Sidhu2018_arxiv}, quantum communications~\cite{Pirandola2020_AOP, Sidhu2021_IET}, and of probabilistic algorithms in computation~\cite{Schaal2020_PRL}. However, quantum mechanics also provides the tools to identify these limits and approach them through the design of practical receivers~\cite{Bae2015_JPA, Burenkov2021_AVS, Burenkov2022_PRL}. Within the framework of optical quantum information science and technology, coherent states of the electromagnetic field play a prominent role as information carriers in terrestrial and space-based quantum networks~\cite{7553489,Sidhu2021_IET}. This is essentially since these states are relatively easy to prepare and control experimentally, and are robust to loss. They are in fact the only pure states that have a classical limit, yet they exhibit fundamental quantum properties when sufficiently attenuated~\cite{ECG,Roy}.

This paper focuses on the quantum-limited detection of coherent states. We present a few explicit and practical schemes for \textit{unambiguous state discrimination} (USD) of coherent states. In particular, instead of considering global bounds, our goal is to design practical receivers that can be realized with linear optics and on-off photodetection. For brevity, we refer to them as \textit{linear receivers}. In some cases, we show that linear receivers nearly saturate the global bounds, which can be computed using the general theory of USD~\cite{Ivanovic1987_PLA, Dieks1988_PLA, Peres1998_JPA, Eldar2003_IEEE, Chefles1998_PLA, Sentis2017_PRL, Nakahira2019_PRA,Sentis2022_Quantum, Bergou2001_PRA, BergouPRL_2012}.

A well-designed quantum receiver combines practicality with high performance, where the latter is quantified through a suitable task-dependent figure of merit. In \textit{ambiguous} state discrimination (ASD), the receiver is designed to always provide an output, with the goal of minimizing the average error probability. The latter is ultimately limited by the Yuen-Kennedy-Lax (YKL) conditions~\cite{Yuen1975-iz} (which reduces to the Helstrom bound for two states~\cite{Helstrom1976}). Alternative figures of merit are preferred for quantum key distribution~\cite{Huttner1995_PRA, Banaszek1999_PLA, VanEnk2002_PRA, Dusek2000_PRA} and quantum digital signatures~\cite{Clarke2012_NC}, where an exact state identification improves the transfer of secure information that cannot be forged or repudiated. This framework is known as USD, where the receiver either identifies the state without error, or outputs an inconclusive result, and the goal is to minimize the average probability of such inconclusive result~\cite{Ivanovic1987_PLA, Dieks1988_PLA}. The ultimate bounds to USD can be computed using the theory of Peres and Terno~\cite{Peres1998_JPA} or that of Sun~\textit{et al.}~\cite{Bergou2001_PRA} and Bergou~\textit{et al.}~\cite{BergouPRL_2012}, which in a sense provide the analogue for the YKL conditions for USD. These bounds can be computed efficiently through semidefinite programming~\cite{Eldar2003_IEEE}, and sometimes allow for an exact analytical solution~\cite{Bergou2001_PRA,BergouPRL_2012}. However, experimentally attaining the bounds, in particular for the discrimination of coherent states, may require high non-linearities that may be un-accessible even with state-of-the-art or near-term future optical technologies. 

There is a wide body of literature devoted to establishing the global bound for USD for different families of quantum states~\cite{Chefles1998_PLA, Sentis2017_PRL, Nakahira2019_PRA, Sentis2022_Quantum, Bergou2001_PRA, BergouPRL_2012}. However, despite these advances, very little is known about practical USD receivers for coherent states, and whether they can achieve the global bounds~\cite{VanEnk2002_PRA, Becerra2013_NC, Izumi2020_PRL, Izumi2021_PRXQ, DiMario2022}, with most of the available works focusing on phase-shift keying. For classical communications using coherent-state modulation, a joint quantum measurement acting on the received coherent-state code word that performs USD achieves the optimal communication capacity allowed by quantum mechanics~\cite{Takeoka2013-al}, known as the Holevo capacity~\cite{Holevo1998-my}. Furthermore, when acting on a finite-length inner code comprised of tensor product of coherent states, the USD measurement can even attain a higher channel capacity---Shannon capacity of the super channel induced by the inner code and the receiver---compared to the optimal ASD measurement that minimizes the average probability of error of choosing between the modulated-received inner code words~\cite{Guha2011-tg, Guha2011-nr, Rosati2016_PRA}. For phase-shift encoded signals, the optimal USD measurement in the limit of small photon numbers consists of mode-wise displacement operations followed by photon-number-resolving detectors~\cite{VanEnk2002_PRA}. For the so-called binary phase-shift-keying (BPSK) alphabet, this scheme is sufficient to attain the optimal USD performance without adaptive pre-detection displacements~\cite{Wittmann2010_PRL}. Post-selecting the measurement result can further reduce the error rate for a fixed probability of inconclusive results~\cite{Wittmann2010_PRA}. For single-mode coherent state constellations with more than two states, there is a substantial gap between the optimal USD performance and non-adaptive receivers~\cite{VanEnk2002_PRA, Becerra2013_NC}. An adaptive receiver for quadrature phase-shift-keying (QPSK) has demonstrated an improvement to correct state identification~\cite{Izumi2020_PRL} in an effort to close this gap.

Here we outline a general theory of USD of coherent states with a receiver that leverages only limited resources, such as multi-mode linear passive optics, phase-space displacement operations, auxiliary vacuum modes, and mode-wise on-off photon detection. We apply our theory to a number of examples of different coherent-state modulations, using as a benchmark the global bounds on USD. The latter are computed explicitly using the theory of Refs.~\cite{Bergou2001_PRA,BergouPRL_2012}. We demonstrate that in some regimes this practical, yet restricted, set of physical operations is typically sufficient to deliver near-optimal performance. This work establishes a theoretical framework to understand and master the design of receivers to enable near-optimal USD of coherent states.


\subsection{Summary of results}
\label{subsec:restuls}

\noindent
Previous works in USD have primarily focused on developing efficient computational methods to determine the global bounds~\cite{Peres1998_JPA, Eldar2003_IEEE, BergouPRL_2012}, providing necessary and sufficient conditions that define optimal measurement schemes~\cite{Chefles1998_PLA, Bergou2001_PRA, Sentis2017_PRL}, or considering specific examples of quantum states to construct measurement operators~\cite{Huttner1996_PRA, Clarke2001_PRA, Nakahira2019_PRA, Izumi2021_PRXQ, Sentis2022_Quantum}.

Here we focus on the design of practical receivers. We introduce a family of receivers based on multimode linear optics and on-off photodetection. A particular strength of these receivers is that they do not require complex adaptive strategies or non-linear optics, which may be challenging to implement. Our work readily addresses the optimal design to minimize the average probability of an inconclusive event. Further, for multimode coherent states, we provide intuitive insights into how the performance of linear receivers depends on the number of modes and on the number of photons detected. This intuition is useful to provide an understanding of how to experimentally realize improvements when using just linear optics. 

We test our receiver design on randomly generated multimode coherent states, showing that linear receivers provide near-optimal performance for a range of coherent states. Non-typical, highly-degenerate constellations of coherent states may necessitate coincidence measurements to achieve USD. We also address an alternative figure of merit, the communication capacity, showing that linear receivers achieve near-optimal performance both in the asymptotic regime and for finite block length.

Table~\ref{tab:usd_receiver_results} provides a high-level summary of the different receiver designs that we introduce in this work, along with their optimal performance bounds, conditions required to saturate these bounds, and how the designs can be adapted to handle general codes. Our class 1 receivers are comprised of vacuum auxiliaries, LOP transformations, and on-off photon detection and are capable of discriminating any full-rank codebook. Our class 2 receivers improve the performance beyond class 1 by using additional mode-wise displacement operations and also extend their applications to codebooks with single degeneracy. Finally, class 3 receivers use the same resources as class 2 but measure detection events across multiple modes. Our results significantly advance the field by clarifying the receiver designs that are both near-optimal for many codes and immediately accessible with current technologies.

\newlength{\thickarrayrulewidth}
\setlength{\thickarrayrulewidth}{2.1\arrayrulewidth}
\renewcommand{\arraystretch}{1.8}
\setlength{\tabcolsep}{8pt}
\begin{table*}[t!]
  \centering
  \begin{tabular}{>{\centering\arraybackslash}m{4cm}|>{\centering\arraybackslash}m{5cm}|>{\centering\arraybackslash}m{7cm}}
    \thickhline
    \textbf{Receiver}\vspace{0pt} & \textbf{Explicit construction}\vspace{0pt} & \textbf{Optimal performance and assumptions} \vspace{0pt}\\
    \hline
    \textbf{Class 1:} single-detection events (Sec.~\ref{subsec:quantitative_bounds})\vspace{0.28cm}	& Fig.~\ref{fig:LOP_scheme}: Vacuum auxiliaries, LOP transformations, and on-off photon detection \vspace{0.0cm}			& Eq.~\eqref{opt0}: requires linearly independent vectors ($\text{rank}(R)=c$) \vspace{0.28cm} \\
    \textbf{Class 2:} single-detection events with displacement (Sec.~\ref{sec:diplace}) \vspace{0.055cm} & Fig.~\ref{fig:LOP_w_dispalcement}: Vacuum auxiliaries, LOP transformations, displacement, and on-off photon detection \vspace{0.05cm}		& Eq.~\eqref{opt1}: vectors at most singularly degenerate ($\text{rank}(R)\geq c-1$) \vspace{0.4cm} \\
    \textbf{Class 3:} double-detection events (Sec.~\ref{sec:2events}) \vspace{0.1cm}				& Fig.~\ref{fig:double_det_receiver}: Vacuum auxiliaries, LOP transformations, displacement, and on-off photon detection \vspace{0.1cm}		& Eq.~\eqref{opt2}: vectors at least doubly degenerate ($\text{rank}(R)\leq c-2$) \vspace{0.3cm}  \\
    Globally optimal receiver (reviewed in App.~\ref{sec:USD_theory} and~\ref{sec:Bergou})				& 
    May require non-linearities and adaptive strategies \vspace{0.03cm}
    & Eq.~\eqref{opt0_PT} or Eq.~\eqref{opt_Bergou} 
    \vspace{0.25cm}
    \\ 
        \thickhline
  \end{tabular}
  \caption{Summary of receiver designs for USD of codes comprised of $c$, $m$-mode coherent states $\ket{\smash{\bm{\alpha}^j}}, j \in [1, c]$. Each class implements on-off photodetection and differs in either the detection strategy or resources used. Note that class 3 receivers can readily be generalized to code words with arbitrary degeneracy by implementing multimode detection events (greater than two). The performance for each class of our linear optical receivers is defined through an optimization problem. We developed a numerical optimizer to address each optimization and derive the optimal solutions. Intuition into the performance of each receiver is provided in section~\ref{subsec:qualitative_bounds}. Note the construction of each receiver class uses linear optics that can be readily implemented, which is in contrast to the globally optimal scenario summarised in the final row.}
  \label{tab:usd_receiver_results}
\end{table*}%
%


\subsection{Outline}
\label{subsec:pa_outline}

\noindent
We begin in Section~\ref{sec:linear_optics} by introducing our notation and the basic theoretical tools to describe our receivers based on linear optics and photodetection. We provide qualitative bounds that introduce key insights into the performance of different detection schemes in Section~\ref{subsec:qualitative_bounds}. Then, in Section~\ref{subsec:quantitative_bounds} we establish a theory to formalize in a quantitative way the problem of USD with limited resources. In Section~\ref{sec:applications}, we discuss applications of linear receivers to different constellations of coherent states, including the pulse-position modulation codes, random codes, and non-typical degenerate codes with single and double degeneracy. These applications demonstrate the strengths of linear receivers when benchmarked against the global bound for USD. Finally, conclusions and open questions are provided in Section~\ref{sec:conc}.


\section{Linear optics in phase space}
\label{sec:linear_optics}

\noindent
Before delving into the details of the USD receivers, we first need to introduce our notation and a few basic elements from the toolbox of linear optics~\cite{Ferraro2005_arxiv}. Consider a collection of $m$ bosonic modes with annihilation and creation operators $\smash{\{a_j, a_j^\dag\}}$, for $j=1,\dots,m$, satisfying the canonical commutation relations, $\smash{[ a_j , a_{j'}^\dag ] = \delta_{jj'}}$. These modes may represent a number of physical degrees of freedom, e.g., polarization, transverse wave vector, time of arrival, orbital angular momentum, as long as they are all degenerate in frequency. A coherent state on mode $j$ is denoted as $|\alpha\rangle_j$ and is characterized by its complex amplitude $\alpha$. We have
\begin{align}
    \ket{\alpha}_j = e^{-\frac{1}{2} |\alpha|^2} \sum_{k=0}^\infty \frac{\alpha^k}{\sqrt{k!}} |k\rangle_j \, ,
\end{align}
where 
\begin{align}
    \ket{k}_j = \frac{1}{\sqrt{k!}} \,  (a_j^\dag)^k |0\rangle
\end{align}
is the Fock state with $k$ photons on mode $j$, and $|0\rangle$ is the vacuum state. A multimode coherent state is the direct product of $m$ coherent states:
\begin{align}\label{c4mxs}
    \ket{\bm{\alpha}} = \bigotimes_{j=1}^m\ket{\alpha_j}_{j} \, .
\end{align}
Such a state is uniquely identified by its amplitude vector $\bm{\alpha}=(\alpha_1, \alpha_2, \dots, \alpha_m)$.
As $|\alpha_j|^2$ is the mean photon number in mode $j$, 
\begin{align}
    n := | \bm{\alpha} |^2 = \sum_{j=1}^m |\alpha_j|^2 
\end{align}
is the total mean photon number in the state. 

The fundamental mathematical structure in quantum mechanics is the scalar product defined in the Hilbert space. The scalar product between two multimode coherent states $|\bm{\alpha}\rangle$ and $|\bm{\beta}\rangle$ reads
\begin{align}
\braket{ \bm{\alpha} \vert \bm{\beta} } = \exp\left[-\frac{1}{2} | \bm{\alpha} |^2 - \frac{1}{2} | \bm{\beta} |^2 + \bm{\alpha}^* \cdot \bm{\beta} \right] \, ,    
\end{align}
where
\begin{align}\label{sp}
\bm{\alpha}^* \cdot \bm{\beta} = 
\sum_{j=1}^m \alpha_j^* \beta_j \, .
\end{align}
is the scalar product between the amplitude vectors. The Hilbert-space scalar product is invariant under the action of general unitary transformations in the Hilbert space and is the central mathematical structure underlying the global bounds on USD~\cite{Peres1998_JPA}. Note that this global bound is known to be achieved by some measurements. However, for generic states, we do not expect any particular form for such optimal measurement.

Since our focus here is on achieving USD using a particular subset of measurements, which includes linear optics and photodetection, the Hilbert-space scalar product may not be the most useful mathematical tool. Therefore, we consider an alternative notion of scalar product that seems more naturally suited to describe quantum mechanics under a restricted set of allowed measurements. To identify this alternative scalar product, we need to consider in more detail the set of unitary linear optics operations, which preserve the total mean photon number. These operations identify the group of Linear Optical Passive (LOP) unitary transformations~\cite{Aniello2006,aaronson2011}.

LOP unitaries map coherent states into coherent states. The components of the amplitude vector transform as follows,
\begin{align}
\alpha_j \to \sum_{k=1}^m U_{jk} \alpha_k    
\end{align}
where $[U_{jk}]$ is a $m \times m$ unitary matrix. Multimode LOP transformations can be implemented physically by combining linear optics elements as beam splitters and phase shifters, and are mathematically described as unitary matrices. Given a unitary matrix description of the LOP transformations, there are known, efficient procedures to simulate it as a network of beam splitters and phase shifters~\cite{ReckZ,Clements,Bell2021_APL}.

Note that the scalar product between amplitude vectors in Eq.~(\ref{sp}) is invariant under the action of LOP unitaries. Therefore, given a pair of $m$-mode coherent states $|\bm{\alpha}\rangle$, $|\bm{\beta}\rangle$, with amplitude vectors $\bm{\alpha}=(\alpha_1, \alpha_2, \dots, \alpha_m)$ and $\bm{\beta}=(\beta_1, \beta_2, \dots, \beta_m)$, we define the \textit{phase-space scalar product} as  
\begin{align}\label{psspdef}
   ( \bm{\alpha}, \bm{\beta} ) 
   = \sum_{j=1}^m \alpha_j^* \beta_j\, .
\end{align}
The phase-space scalar product will play in our analysis a similar role played by the Hilbert-space scalar product in the theory of USD developed by Peres and Terno~\cite{Peres1998_JPA}, and it will guide us in designing our linear receivers for USD of coherent states.

In the rest of the paper, we will consider examples of codes comprising $c \geq 2$ $m$-mode coherent states, identified by $c$ amplitude vectors $\bm{\alpha}^1$, $\bm{\alpha}^2$, $\dots$, $\bm{\alpha}^c$. We can arrange these vectors into a rectangular matrix with $c$ rows and $m$ columns, 
\begin{align}\label{Rank0}
    R = \left( 
    \begin{array}{c}
    \bm{\alpha}^1 \\
    \bm{\alpha}^2 \\
    \vdots \\
    \bm{\alpha}^c 
    \end{array}    
    \right) 
    = \left( 
    \begin{array}{cccc}
    \alpha^1_1 & \alpha^1_2 & \dots & \alpha^1_m \\
    \alpha^2_1 & \alpha^2_2 & \dots & \alpha^2_m \\
    \vdots & \vdots & \vdots & \vdots\\
    \alpha^c_1 & \alpha^c_2 & \dots & \alpha^c_m 
    \end{array}    
    \right) \, .
\end{align}
It is well known that pure states can be unambiguously discriminated if and only if they are linearly independent~\cite{Chefles1998_PLA}. In our setting, where we only allow for limited resources, this condition needs to be modified. As a matter of fact, we need to look at the notion of linear independence in phase space, not in the Hilbert space. This alternative notion of linear independence, which we simply call \textit{phase-space linear independence}, is strictly related to the phase-space scalar product introduced above. Therefore, we will say that the given coherent states are linear independent in phase space if their associated amplitude vectors are linear independent, i.e., if the matrix $R$ in Eq.~(\ref{Rank0}) has maximum rank. Note that this is possible only if $c \leq m$. Below we describe a scheme for USD of coherent states that works when the associated $R$-matrix is full rank.


In general, our receivers will also exploit a number of auxiliary modes. We will then introduce $m'$ additional optical modes (as we see below, it is sufficient to use $m'\leq m$ auxiliary modes), characterized by the canonical operators $\{ a_{m+j}$, $a_{m+j}^\dag \}$, for $j=1$ to $m'$. The totality of $m+m'$ will be mixed at a LOP unitary over $m+m'$ modes, which is represented by a unitary matrix of size $m+m'$. If all modes are initially prepared in a coherent state, with amplitude vector $\bm{\alpha} = (\alpha_1 , \dots, \alpha_m, \alpha_{m+1} , \dots \alpha_{m+m'})$, then the output is also a coherent state, with amplitude vector $\bm{\beta} = (\beta_1 , \dots, \beta_m, \beta_{m+1} , \dots \beta_{m+m'})$,
where the $i$th output amplitude reads
\begin{align}
\beta_i = \sum_{j=1}^{m+m'} U_{ij} \alpha_j = \sum_{j=1}^{m} M_{ij} \alpha_j + \sum_{j=m+1}^{m+m'} N_{i j} \alpha_{j} \, ,
\end{align}
where $M$ and $N$ are submatrices of $U$, with $M_{ij}= U_{ij}$ for $j = 1,\dots, m$, and $N_{ij}= U_{i j}$ for $j = m+1, \dots, m+m'$. For our applications, we are often interested in the case where the $m'$ ancillary modes are prepared in the vacuum state, i.e., $\alpha_{m+j}=0$. In this setting, we simply have 
\begin{align}
\beta_i = \sum_{j=1}^{m} M_{ij} \alpha_j \, .
\end{align}

When the output coherent state is measured by mode-wise photodetection, for any $i$ we will have a certain probability of detecting a photon in the $i$th mode. This probability is readily computed as
\begin{align} \label{prob-det}
P_i(\bm{\alpha}, U)
= 1 - \exp{ \left[ - | \bm{M}_i \cdot \bm{\alpha} |^2 \right] } \, ,
\end{align}
where 
\begin{align}
\bm{M}_i = ( M_{i1} , M_{i2} , \dots , M_{im})    
\end{align}
and
\begin{align}
 \bm{M}_i \cdot \bm{\alpha} = \sum_{j=1}^m M_{ij} \alpha_j \, .   
\end{align}
The photon detection probability in Eq.~\eqref{prob-det} is the key quantity to characterize our linear receivers, as we discuss in Section~\ref{subsec:quantitative_bounds}.

Finally, we recall that the displacement operator, denoted as $D(\gamma)$, maps a coherent state into a coherent state with a displaced amplitude:
\begin{align}
    D(\gamma) |\alpha\rangle = |\alpha + \gamma\rangle \, .
\end{align}
In the $m$-mode case, the displacement operator is identified by a complex displacement vector, $\bm{\gamma} = (\gamma_1, \gamma_2, \dots, \gamma_{m})$, such that
\begin{align}
    D( \bm{\gamma} ) |\bm{\alpha}\rangle = |\bm{\alpha} + \bm{\gamma} \rangle \, ,
\end{align}
where $\bm{\alpha} + \bm{\gamma} = (\alpha_1 + \gamma_1, \alpha_2 + \gamma_2, \dots, \alpha_{m} + \gamma_{m})$
and $D( \bm{\gamma} ) = D(\gamma_1) \otimes D(\gamma_2) \otimes \cdots D(\gamma_{m})$.


\section{Qualitative analysis of linear receivers for USD of coherent states}
\label{subsec:qualitative_bounds}

\noindent
This work addresses the following general question: Given a collection of $c$, $m$-mode coherent states $\ket{\smash{\bm{\alpha}^j}}$, with amplitude vectors $\bm{\alpha}^j = ( \alpha^j_1 , \alpha^j_2 , \dots , \alpha^j_m)$ and prior probabilities $p^j$, is it possible to discriminate them unambiguously using linear optics, $m'$ auxiliary vacuum modes, and mode-wise photodetection? In this Section, we provide a qualitative answer to this question. These qualitative results will give some insight into the numerical analysis discussed later in the paper.

Consider a set of $m$-mode coherent states with $n$ mean photon number. If we use $m'$ ancillary vacuum modes, then the mean photon number per mode is $\nu = n/(m+m')$. These states are processed through a LOP unitary, which preserves the photon number, and then measured by mode-wise photodetection. On average, the probability of a single photodetection event in a given mode is expected to be about
\begin{align}
\label{onedet}
1 - e^{-f\nu}    \, .
\end{align}
where the factor $f$ depends on the details of the input states and LOP transformation applied. For weak signals, it is unlikely to observe more than one photodetection event. Therefore, we cannot discriminate more than $m+m'$ states, i.e., no more than the number of signal plus auxiliary modes, and the receiver yields an inconclusive result every time no photon is detected. This happens with probability 
\begin{align}\label{taylor1}
P_0 = e^{- f\nu (m+m')} = e^{-fn} = 1 - fn + O(n^2)\, .
\end{align}
Therefore, we expect the probability of an inconclusive event to decrease linearly with $n$ for $n \ll 1$. This probability can be further decreased if instead of vacuum auxiliary modes we use auxiliary coherent states. In fact, this would increase the mean photon number from $n$ to $n'>n$. This strategy is equivalent to introducing phase-space displacement in the pool of allowed resources~\cite{Sidhu2020_arxiv}. In Section \ref{subsec:random_codes}, see Fig.~\ref{fig:Random_codebooks}(a) in particular, we will show that this qualitative behaviour is confirmed quantitatively, both with and without phase-space displacement.

For brighter coherent states, or when the first order term in $n$ vanishes, we need to consider joint detection events on pairs of modes. In this case, the number of distinct outcomes is
\begin{align}
{    m+m' \choose 2 } \, ,
\end{align}
which gives the maximum number of states that can be unambiguously discriminated by observing coincidence detection on two modes. From Eq.~(\ref{onedet}), and recalling that a multimode coherent state is a product state, see Eq.~(\ref{c4mxs}), we obtain the probability of a coincidence
\begin{align}
(1 - e^{-f \nu })^2 \, .
\end{align}
By ignoring joint detection events in more than two modes, the probability of obtaining an inconclusive result is
\begin{align}\label{P0-double}
P_0 & = e^{- f \nu (m+m')} \\
& \quad + (m+m')e^{- f \nu (m+m'-1)} \left( 1 - e^{- f \nu } \right) \nonumber \\
& = 1 - {m+m' \choose 2} f^2 \nu^2 + O(\nu^3) \,.
\end{align}
Recalling that $\nu = n/(m+m')$, we obtain
\begin{align}
P_0 & = 1 - \frac{1}{2}\frac{m+m'-1}{m+m'} f^2 n^2 + O(n^3) \\
& \simeq 1 - \frac{1}{2} f^2 n^2 + O(n^3) \, .
\end{align}
In this case, the probability of the inconclusive event decreases quadratically with $n$ and, for $m+m'$ large enough, is essentially independent of the number of modes and ancillas. As for the previous case, using coherent states ancillas instead of vacuum ones, we may further decrease $P_0$, as this is equivalent to replacing $n$ with $n'>n$. This qualitative result will be confirmed by the quantitative analysis of Section~\ref{subsec:degenerate_codes-2}, see Fig.~\ref{fig:double_degenerate_codes}(b) in particular, where we will discuss examples of degenerate codes where the linear term in Eq.~(\ref{taylor1}) vanishes.

In principle, a similar qualitative analysis can be applied to USD schemes based on an arbitrary number of joint detection events. We naturally expect that events of joint detection of $n_d$ photons will contribute to $P_0$ with a term proportional to $n^{n_d}$. However, in this work, we only discuss receivers based on single and double-detection events. In the following Section, we will focus on photodetection events on at most one mode, and provide quantitative analysis and explicit USD schemes. Later in Section~\ref{subsec:degenerate_codes-2} we will consider an example of USD based on joint detection on two modes.


\section{Explicit schemes for USD of coherent states}
\label{subsec:quantitative_bounds}

\noindent
Below we present three schemes for linear receivers. The first two, in Sections \ref{sec:1event} and \ref{sec:diplace}, exploit a single detection event to achieve USD of coherent states. The third scheme in Section \ref{sec:2events} exploits double detection events. These schemes are not necessarily optimal for the given resources, however, we show that they are optimal or near-optimal in a number of instances.


\subsection{Single detection events}
\label{sec:1event}

\begin{figure}[t!]
\centering
\includegraphics[width=0.9\linewidth]{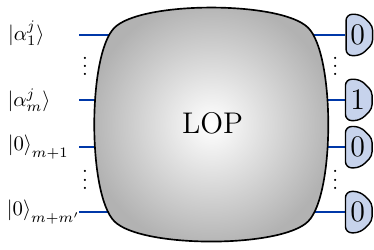}
\caption{Linear receiver design, based on LOP unitary transformations and mode-wise on-off photodetection for USD of coherent states. The input coherent states are distributed across $m$ input modes with $m'$ vacuum auxiliary modes. Note that the LOP unitaries can be efficiently implemented through a network of beam splitters and phase shifters~\cite{ReckZ,Clements,Bell2021_APL}.}
\label{fig:LOP_scheme}
\end{figure}

\noindent
We now introduce an explicit scheme for USD of multimode coherent states, based on single photo-detection events. Consider a set of $c$ coherent states over $m$ modes, identified by the amplitude vectors $\bm{\alpha}^1$, $\bm{\alpha}^2$, $\dots$, $\bm{\alpha}^c$, where $\bm{\alpha}^j = (\alpha_1^j, \alpha_2^j, \dots, \alpha_m^j)$. In order to discriminate these states, we first introduce $m'$ auxiliary vacuum modes, then mix them on a LOP transformation identified by the $(m+m') \times (m+m')$ unitary matrix $U$, and finally apply mode-wise on-off photodetection. This setup is shown schematically in Fig.~\ref{fig:LOP_scheme}. As recalled in Section \ref{sec:linear_optics}, the probability that, given the state $|\bm{\alpha}^j \rangle$ in input, a photon is detected on mode $i$ in output, is
\begin{align} 
P_i(\bm{\alpha}^j, U)
= 1 - \exp{ \left[ - | \bm{M}_i \cdot \bm{\alpha}^j |^2 \right] } \, ,
\end{align}
where the vectors $\bm{M}_i$ are determined by the elements of the matrix $U$ as discussed in Section \ref{sec:linear_optics}.

If we want to achieve USD using the information obtained from a single photodetection event in one of the output modes, then it is obvious that we need to impose that the input states are in one-to-one correspondence with the output modes. Without loss of generality, this means that for the $j$th state in input all the output modes are in the vacuum except the $j$th mode. That is, the matrix $U$ needs to be chosen in such a way that
\begin{align} 
P_j(\bm{\alpha}^j, U) & \geq 0 \, , \\
P_i(\bm{\alpha}^j, U) & = 0  \, \, \text{for} \, \, i \neq j \, .
\end{align}
Expressed in terms of the vectors $\bm{M}_i$, this condition reads
\begin{align}
\label{indepcond}
\bm{M}_i \cdot \bm{\alpha}^j = \delta_{ij} \bm{M}_i \cdot \bm{\alpha}^j \, .
\end{align}
We note that a non-zero matrix $M$ that solves these equations exists if and only if the amplitude vectors are linearly independent, i.e., the matrix $R$ in Eq.~(\ref{Rank0}) is full rank~\footnote{Equation~(\ref{indepcond}) emerges in our discussion of linear USD receivers. This condition plays the same role as Eq.~(\ref{vec_overlap}) which appears in the general theory of USD, see for example the work of Peres and Terno~\cite{Peres1998_JPA}. Notably, the two equations are formally equivalent upon replacing the Hilbert-space scalar product with the phase-space scalar product introduced in Section \ref{sec:linear_optics}. In this sense, our theory is a modification of the approach of Peres and Terno, but it is defined in phase space instead of the Hilbert space.}.

The conditions in Eq.~(\ref{indepcond}) ensure that the detection of a photon in output mode $j$ unambiguously identifies the coherent state $|\bm{\alpha}^j\rangle$ in input. Otherwise, no conclusion can be drawn in the event that no photon is detected in any of the output modes. This is the inconclusive event. If the input coherent state $|\bm{\alpha}^j\rangle$ has probability $p^j$, then the average probability of the inconclusive event is
\begin{align}
P_0 = \sum_{j=1}^c p^j \exp{ [- | \bm{M}_j \cdot \bm{\alpha}^j |^2 ]} \, .
\end{align}
Our goal is to design an explicit receiver that minimizes this probability. This corresponds to performing an optimization on the $c \times m$ matrix $M$, keeping in mind that the latter is by construction a submatrix of a larger unitary matrix. This latter condition is expressed by the matrix inequality 
\begin{align}
 M^\dag M \leq I \, ,
\end{align}
where $I$ is the identity matrix (see Appendix~\ref{sec:Mconstraint} for a derivation of this condition). In solving the constrained optimization, we parameterize the elements of the matrix $M$ as 
\begin{align}
M_{ij} = \sqrt{k_i} \, v_{ij} \, ,    
\end{align}
where $k_i \geq 0$ are $c$ non-zero coefficients, and $v_{ij}$ are the components of $c$ unit vectors
\begin{align}
\bm{v}_i = (v_{i1}, \dots, v_{im}) \, ,
\end{align}
for $i=1,\dots, c$. The unit vectors $\bm{v}_i$ are proportional to the vectors $\bm{M}_i$, therefore condition (\ref{indepcond}) becomes
\begin{align}
\bm{v}_i \cdot \bm{\alpha}^j = \delta_{ij} \bm{v}_i \cdot \bm{\alpha}^j \, .
\end{align}


In conclusion, using this parameterization, the optimal linear receiver is determined by solving the constrained optimization:
\begin{mini} 
{k_1, \dots, k_c}{\sum_{j=1}^c p^j \exp{\left[ - k_j \left| \bm{v}_j \cdot \bm{\alpha}^j \right|^2 \right]} \, ,}
{}{}
\addConstraint{\bm{v}_i \cdot \bm{\alpha}^j}{ = \delta_{ij} \bm{v}_i \cdot \bm{\alpha}^j}
\addConstraint{M_{ij}}{=\sqrt{ k_i} \, v_{ij}}
\addConstraint{ M^\dag M  }{\leq I} \, .
\label{opt0}
\end{mini}
The minimal value represents the minimum probability of obtaining an inconclusive event for distinguishing the given set of coherent states using only vacuum auxiliary modes, LOP unitaries, and on-off photodetection. This value can be compared to the global bound obtained when general operations and measurements are allowed, which can be computed analytically or numerically using results already available in literature~\cite{BergouPRL_2012}.

Finally, we remark that the optimization problem (\ref{opt0}) is not explicitly dependent on the number of auxiliary modes $m'$. However, once an optimal form for the matrix $M$ is obtained, one needs to find a unitary matrix that extends it. In general, such a matrix exists only if $m'$ is chosen sufficiently large. However, one can assume $m' \leq m$ without loss of generality. An explicit construction is given in Appendix~\ref{sec:Mconstraint}.


\subsection{Phase-space displacement improves USD}
\label{sec:diplace}

\begin{figure}[t!]
\centering
\includegraphics[width=\linewidth]{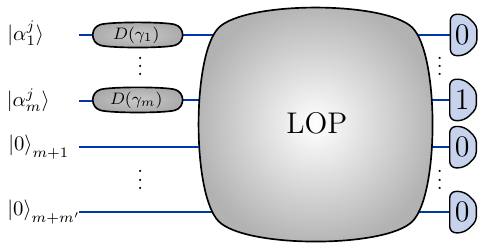}
\caption{Linear receiver with improved performance, employing phase-space displacement operations.}
\label{fig:LOP_w_dispalcement}
\end{figure}

\noindent
Better USD schemes, i.e., with a lower probability of the inconclusive event, can be obtained by enlarging the set of allowed resources. Here we describe a USD scheme obtained by adding the operation of phase-space displacement. We consider a setup where multimode displacement is first applied to the input modes, followed by the LOP unitary and mode-wise photodetection. This is shown in Fig.~\ref{fig:LOP_w_dispalcement}. 

The optimal setup is then obtained by minimizing the probability of the inconclusive event, where now there are $m$ additional complex degrees of freedom, corresponding to the components of the displacement vector $\bm{\gamma}$:
\begin{mini} 
{k_1, \dots, k_c, \bm{\gamma}}{\sum_{j=1}^c p^j \exp{\left[ - k_j \left| \bm{v}_j \cdot \left( \bm{\alpha}^j + \bm{\gamma} \right) \right|^2 \right]} \, ,}
{}{}
\addConstraint{\bm{v}_i \cdot ( \bm{\alpha}^j + \bm{\gamma} )}{= \delta_{ij} \bm{v}_i \cdot \left( \bm{\alpha}^j + \bm{\gamma} \right)}
\addConstraint{M_{ij}}{=\sqrt{ k_i} \, v_{ij}}
\addConstraint{M^\dag M}{\leq I} \, .
\label{opt1}
\end{mini}
The examples of Section \ref{sec:applications} will show that the introduction of phase-space displacement may improve the performance of USD substantially.

As we noted above, the constraint
\begin{align}
    \bm{v}_i \cdot ( \bm{\alpha}^j + \bm{\gamma} ) = \delta_{ij} \bm{v}_i \cdot \left( \bm{\alpha}^j + \bm{\gamma} \right)
\end{align}
can be satisfied if and only if the displaced amplitude vectors $\bm{\beta}^j = \bm{\alpha}^j + \bm{\gamma}$ are linearly independent. The use of displacement operations has the added value of casting a linearly-dependent system with single degeneracy \footnote{We call \textit{degenerate code} a set of coherent states whose amplitude vectors are linearly dependent in phase space. Therefore, the associated $R$ matrix is not full rank. We say that the code has single degeneracy if the rank of the $R$ matrix is one unit below the maximum value, and has double degeneracy if it is two units below the maximum value.} into an independent one. This extends the range of coherent states that can be discriminated unambiguously. To see this, consider a linearly dependent set of coherent states with $c\leq m$, and rank-deficient $R$ matrix with $\mathrm{rank}(R)=c-1$. The matrix can be made full rank by adding to each row a linear independent vector $\bm{\gamma}$. The new matrix,
\begin{align}
    R_1 = \left( 
    \begin{array}{c}
    \bm{\alpha}^1 + \bm{\gamma} \\
    \bm{\alpha}^2 + \bm{\gamma}\\
    \vdots \\
    \bm{\alpha}^c + \bm{\gamma}
    \end{array}    
    \right) \, ,
\end{align}
has full rank and represents the set of displaced coherent states.

Displacement also comes to the rescue when the number of states is larger than the number of modes, in the case $c=m+1$ and $\mathrm{rank}(R)=c-1$. To make the matrix full rank, we first add one auxiliary mode, then displace the $m+1$ modes by $(\bm{\gamma},\gamma_{m+1})$. 
The new matrix reads
\begin{align}
    R_2 = \left( 
    \begin{array}{cc}
    \bm{\alpha}^1 + \bm{\gamma} & \gamma_{m+1} \\
    \bm{\alpha}^2 + \bm{\gamma}& \gamma_{m+1} \\
    \vdots & \vdots \\
    \bm{\alpha}^c + \bm{\gamma}& \gamma_{m+1}  
    \end{array}    
    \right) \, ,
\end{align}
and has rank $c$ for suitable choices of $(\bm{\gamma},\gamma_{m+1})$. An explicit example of this method is presented in Section \ref{subsec:phasemod}.


\subsection{Double detection events}
\label{sec:2events}

\noindent
When the amplitude vectors are not linearly independent and the matrix $R$ has multiple degeneracies, auxiliary modes and phase-space displacements are not sufficient to make the matrix full rank. Note that the only way to do that is to use a state-dependent displacement, which would imply some prior knowledge of the code word.

\begin{figure}[t!]
\centering
\includegraphics[width=0.9\linewidth]{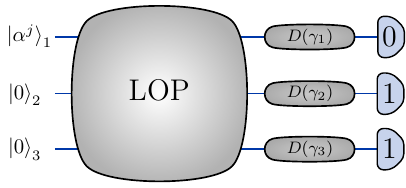}
\caption{Linear receiver employing double detection events to unambiguously distinguish doubly degenerate codes with $m=3$, $c=1$. }
\label{fig:double_det_receiver}
\end{figure}

To bypass this problem, here we focus on an alternative approach based on multiple detection events. As an example, we consider the simplest family of doubly degenerate codes, which is obtained for $c=3$ and $m=1$, i.e., a code of three coherent states over one mode: $|\alpha^1\rangle$, $|\alpha^2\rangle$, $|\alpha^3\rangle$. To discriminate these states, we introduce an explicit linear receiver design that makes use of two auxiliary vacuum modes. The three modes (one signal and two auxiliary modes) are first mixed in LOP unitary, then displaced by $\gamma_1$, $\gamma_2$, $\gamma_3$, and finally detected as schematically shown in Fig.~\ref{fig:double_det_receiver}.
Unlike the receivers of Sections~\ref{sec:1event}, \ref{sec:diplace}, here two joint detection events unambiguously determine the input state. Without loss of generality, we require that input state $|\alpha_1\rangle$ yields the vacuum in the first output mode, whereas the other two output modes both have a non-zero probability of photon detection. Similarly, we require that input state $|\alpha_2\rangle$ yields the vacuum state on the second output mode, and state $|\alpha_3\rangle$ yields the vacuum on the third output mode.

To write this condition explicitly, consider the unitary matrix $U_{ki}$ that represents the LOP transformation. Then the amplitude on the output mode $k$, given the input state $|\alpha^j\rangle$ is
\begin{align}
    \zeta_k^j = U_{k1} \alpha^j + \gamma_k \, .
\end{align}
We require that the output amplitude vanishes for $j=k$, i.e.,
\begin{align}
    0 = U_{j1} \alpha^j + \gamma_j \, ,
\end{align}
from which we obtain three constraints:
\begin{align}\label{condo2}
    \gamma_j = - U_{j1} \alpha^j \, , 
\end{align}
for $j=1,2,3$.
Given the input state $|\alpha^1\rangle$, the probability of obtaining two joint photodetection events on modes $2$ and $3$ is
\begin{align}
    P(2,3|1) 
    & = \left( 1 - e^{-|\zeta_2^1|^2} \right)
        \left( 1 - e^{-|\zeta_3^1|^2} \right) \\
    & = \left( 1 - e^{-|U_{21} \alpha^1 + \gamma_2|^2} \right)
        \left( 1 - e^{-|U_{31} \alpha^1 + \gamma_3|^2} \right) \\
    & = \left( 1 - e^{-|U_{21}|^2 |\alpha^1 - \alpha^2 |^2} \right)
        \left( 1 - e^{-|U_{31}|^2 |\alpha^1 - \alpha^3|^2} \right) \, ,
\end{align}
where in the last equation we have used condition (\ref{condo2}). Note that $P(2,3|1)$ is the probability of identifying the input state $|\alpha^1\rangle$, as we use double photodetection events to unambiguously discriminate the input states. From this expression, given the prior probability $p^j$ for coherent state $|\alpha^j\rangle$, we compute the average probability of obtaining an inconclusive event:
\begin{align}
    P_0 & = 1 - p^1 P(2,3|1) - p^2 P(3,1|2) - p^3 P(1,2|3) \\
    & = 1 
    - p^1 \left( 1 - e^{-|U_{21}|^2 |\alpha^1 - \alpha^2 |^2} \right)
    \left( 1 - e^{-|U_{31}|^2 |\alpha^1 - \alpha^3|^2} \right) \nonumber \\
    & \phantom{=} - p^2 \left( 1 - e^{-|U_{31}|^2 |\alpha^2 - \alpha^3 |^2} \right)
    \left( 1 - e^{-|U_{11}|^2 |\alpha^2 - \alpha^1|^2} \right) \nonumber \\
    & \phantom{=} - p^3 \left( 1 - e^{-|U_{11}|^2 |\alpha^3 - \alpha^1 |^2} \right)
    \left( 1 - e^{-|U_{21}|^2 |\alpha^3 - \alpha^2|^2} \right) \, .
\end{align}
The optimal performance of linear receivers based on double detection events is then obtained by minimizing this expression for $P_0$ under the constraint that the matrix $U$ is unitary:
\begin{mini} 
{U_{11}, U_{12}, U_{13}}{P_0 \, , \hspace{3.3cm}}
{}{}
\addConstraint{|U_{11}|^2 + |U_{21}|^2 + |U_{31}|^2}{=1} \, .
\label{opt2}
\end{mini}
This constrained optimization problem can be solved using Lagrange multipliers. Using the same setup, we can generalize this approach to codes with higher degeneracies.

%


\section{Examples}
\label{sec:applications}

\noindent
In this Section, we explore with a few examples the performance of our linear optical USD receivers relative to the global bound, for different coherent states and figures of merits. We illustrate that, despite its simplicity, decoders constructed using only linear components and on-off photodetection can generate near-optimal USD: this is observed for randomly generated coherent states, where phase-space displacement is necessary to achieve near-optimal performances. The situation appears to be different for non-typical coherent states with multiple degeneracies, in which case we need to exploit double-detection events.


%
\begin{figure*}[t!]
    \centering
    \includegraphics[width=\linewidth]{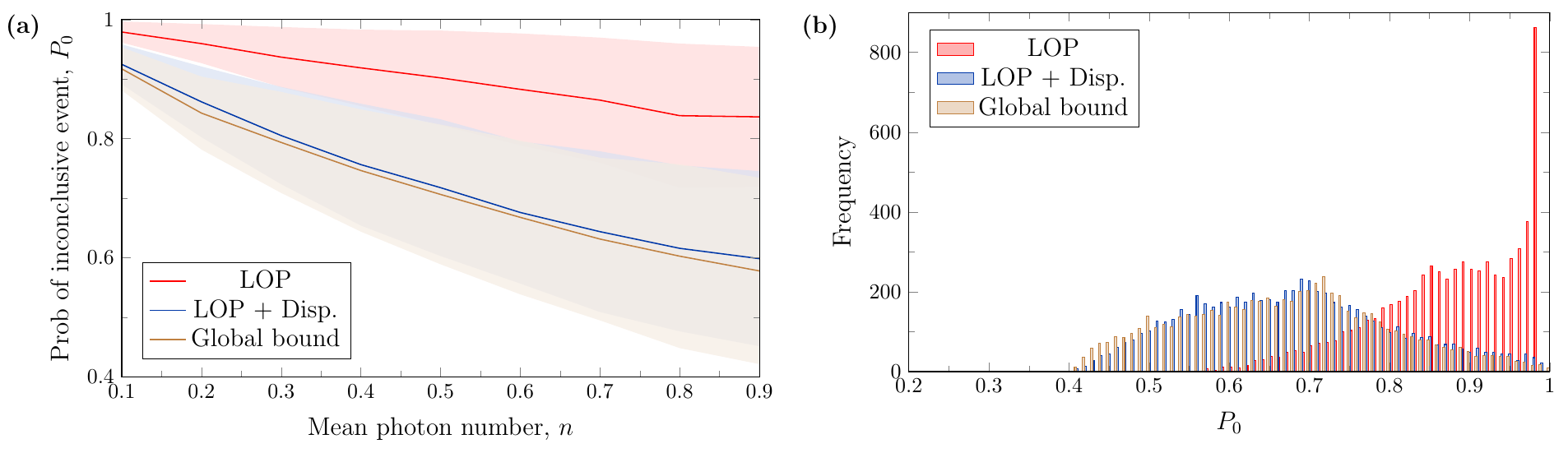}
    \caption{Benchmarked performance of the linear receivers with displacement (\ref{opt1}) and without displacement (\ref{opt0}), sampled on random codes. (a) Probability of inconclusive event $P_0$ plotted vs the photon number, obtained by sampling over $N=500$ random codes for each $n$. Shaded regions account for data within one standard deviation. The resolution for the mean photon number is $\Delta n =0.1$. (b) Frequency distribution of the inconclusive event probability, obtained from $N=6600$ random samples with $n=0.6$. The global bound is computed semi-analytically using the method of Bergou \emph{et al.}~\cite{BergouPRL_2012}.}
    \label{fig:Random_codebooks}
\end{figure*}%

\subsection{Discrimination of PPM codes and equivalent codes}
\label{subsec:PPM}

\noindent
A set of orthogonal vectors can always be discriminated perfectly. Since coherent states with finite energy are never orthogonal, they cannot be perfectly discriminated (though their discrimination can be improved by increasing their distance in phase space). In this Section, we explore a particular family of coherent states that share some formal features with orthogonal states. A pulse-position modulation (PPM) code is a set of $c=m$ coherent states over $m$ modes: 
\begin{align}
\begin{split}
\ket{\smash{\bm{\alpha}^1}} & = \ket{\alpha , 0, \dots , 0} \, ,\\
\ket{\smash{\bm{\alpha}^2}} & = \ket{0 , \alpha , \dots , 0} \, ,\\
 & \vdots \\
\ket{\smash{\bm{\alpha}^m}} & = \ket{0 , 0, \dots , \alpha}  \, ,
\end{split}
\label{eqn:PPM_cd}
\end{align}
such that the matrix $R = \alpha I$ is a full-rank. Note that these coherent states are mutually orthogonal with respect to the phase-space scalar product (defined in Eq.~(\ref{psspdef})), with Gram matrix $(\bm{\alpha}^i,\bm{\alpha}^j) = |\alpha|^2 \delta_{ij}$. The PPM code is often discussed in the context of quantum communications \cite{Guha2011-tg,Guha2011-nr,enigma}. In general, it represents a case study for quantum state discrimination and hypothesis testing, see for example Ref.~\cite{Skotiniotis2018}. 

In general, there is a gap between the global bound and the performance of linear receivers. However, for a PPM code, it is easy to show that the gap vanishes as the PPM code can be optimally discriminated against using on-off photodetection only. In fact, to unambiguously discriminate the states in the PPM code, it is sufficient to apply mode-wise photodetection. If a photon is detected on mode $j$, then we know the input state necessarily was $|\smash{ \bm{\alpha}^j}\rangle$ with no ambiguity. The inconclusive event is when no click is recorded, which happens with probability $\smash{P_0 = e^{-|\alpha|^2}}$. 
It is well known that this inconclusive probability saturates the global bound. To show this, one can apply the results of Bergou \emph{et al.}~in Ref.~\cite{BergouPRL_2012} (reviewed in Appendix~\ref{sec:Bergou}). 

The optimality of linear receivers extends to a larger class of codes beyond PPM. First note that, since the phase-space scalar product is invariant under LOP unitaries, it immediately follows that LOP unitaries and on-off photodetection are sufficient to optimally discriminate any set of coherent states that has the same Gram matrix as the PPM code
\begin{equation}
( \bm{\alpha}^i , \bm{\alpha}^j ) = |\alpha|^2 \delta_{ij} \, ,
\end{equation}
If the Gram matrix is not diagonal, we may try to make it diagonal by applying a phase-space displacement $\bm{\alpha}^j \to \bm{\beta}^j = \bm{\alpha}^j + \bm{\gamma}$, which changes the Gram matrix into
\begin{align}
( \bm{\beta}^i , \bm{\beta}^j ) & =
( \bm{\alpha}^i , \bm{\alpha}^j )  
+ ( \bm{\alpha}^i , \bm{\gamma} )
+ ( \bm{\gamma} , \bm{\alpha}^j )  
+ ( \bm{\gamma} , \bm{\gamma} )  \, .
\end{align}
Therefore, the code can be optimally discriminated with LOP unitaries, displacements, and photodetection, if there exists $\bm{\gamma}$ and $\tau>0$ such that
\begin{align}
( \bm{\alpha}^i , \bm{\alpha}^j )  
+ ( \bm{\alpha}^i , \bm{\gamma} )
+ ( \bm{\gamma} , \bm{\alpha}^j )  
+ ( \bm{\gamma} , \bm{\gamma} )  =  \tau \delta_{ij} \, .
\end{align}
This is a system of $c^2$ real equations and $2 m + 1$ real unknowns (the components of the complex displacement vector $\bm{\gamma}$ and $\tau$). Therefore, in general, we expect this system of equations to have solutions if $c^2 \leq 2 m + 1$.

As we have seen above, sometimes the use of an auxiliary mode can improve the effectiveness of linear receivers. In fact, adding an auxiliary mode allows us to introduce one additional real degree of freedom, i.e., $|\gamma_{m+1}|^2$. The new system of equations reads
\begin{align}
( \bm{\alpha}^i , \bm{\alpha}^j )  
+ ( \bm{\alpha}^i , \bm{\gamma} )
+ ( \bm{\gamma} , \bm{\alpha}^j )  
+ ( \bm{\gamma} , \bm{\gamma} ) + |\gamma_{m+1}|^2 =  \tau \delta_{ij} \, ,
\label{eqn:sys_eqs}
\end{align}
and comprises $c^2$ equations and $2m+2$ unknowns. Therefore, we expect this to generally admit a solution if $c^2 \leq 2m + 2$. Note that there is no benefit in adding more than one auxiliary mode~\footnote{If we add $k$ auxiliary modes we obtain the following system of equations:
$( \bm{\alpha}^i , \bm{\alpha}^j )  
+ ( \bm{\alpha}^i , \bm{\gamma} )
+ ( \bm{\gamma} , \bm{\alpha}^j )  
+ ( \bm{\gamma} , \bm{\gamma} ) + \sum_{j=1}^k |\gamma_{m+j}|^2 =  \tau \delta_{ij}$.
Defining the real parameter $\smash{\Gamma = \sum_{j=1}^k |\gamma_{m+j}|^2}$, this system of equations is equivalent to having a single auxiliary mode.}.


\subsection{Dual of the PPM code}
\label{subsec:dual}

\noindent
An example of code that can be reduced to PPM by applying phase space displacement is the following
\begin{align}
\begin{split}
\ket{\smash{\bm{\alpha}^1}} & = \ket{0 , \alpha, \alpha \dots, \alpha} \, ,\\
\ket{\smash{\bm{\alpha}^2}} & = \ket{\alpha, 0, \alpha, \dots, \alpha} \, ,\\
\vdots & \\
\ket{\bm{\alpha}^m} & = \ket{\alpha, \alpha, \dots , \alpha, 0}  \, ,
\end{split}
\label{eqn:dualPPM}
\end{align}
where the state $\ket{\bm{\alpha}^j}$ has the vacuum on mode $j$ and a coherent state of given amplitude $\alpha$ in the other modes. Note that this code is mapped into a PPM code by displacing each mode by $-\alpha$. Therefore, the linear receiver is optimal and saturates the global bound on the inconclusive event probability, $\smash{P_0 = e^{-|\alpha|^2}}$.


\subsection{Random codes}
\label{subsec:random_codes}

\noindent
In this Section, we illustrate the performance of linear receivers on random states. Random codes are particularly beneficial in benchmarking the performance of receivers~\cite{Nazer2008_ETT}. As an example, we consider $c=3$ coherent states over $m=3$ modes. We sample their amplitudes $(\alpha_1, \alpha_2, \alpha_3)$ uniformly from a sphere of radius $\sqrt{n}$, where $n = \sum_i |\alpha_i|^2$, where for simplicity we restrict to real-valued amplitudes. Figure~\ref{fig:Random_codebooks}(a) illustrates the minimized inconclusive probability $P_0$ achieved from our optimized framework without (\ref{opt0}) and with (\ref{opt1}) displacement. Notice that the dependence of $P_0$ on the average photon number is compatible with the exponential law obtained in the qualitative analysis of Section \ref{subsec:qualitative_bounds}. Without displacement, the linear receivers perform poorly. However, when equipped with displacement the performance improves significantly and nearly matches the global bound. To see this more clearly, Fig.~\ref{fig:Random_codebooks}(b) illustrates the statistical distributions of $P_0$ for $n=0.6$ with $N=6600$ random codes. The distribution of the inconclusive probability for the linear receivers with displacement closely matches the global bound, with small variations at smaller values of $P_0$. The global bound has been computed using the method of Ref.~\cite{BergouPRL_2012}. This result reassuringly demonstrates that practical receivers based on linear optics are near-optimal and sufficient to match the global bound, as long as phase-space displacement operations are accessible.


\subsection{Codes with single degeneracy}
\label{subsec:degenerate_codes-1}

\begin{figure*}[t!]
\centering
\includegraphics[width=0.98\linewidth]{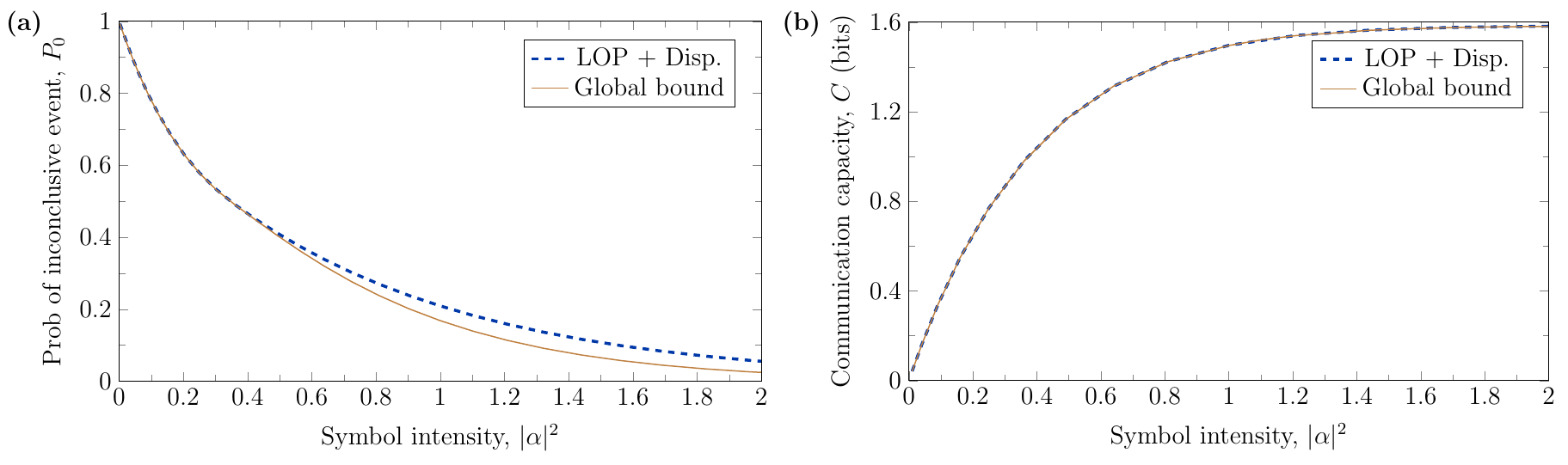}
\caption{Benchmarked performance of linear receivers for the two-mode code with single degeneracy from Eq.~(\ref{eq:Guha2011}). (a) Probability of inconclusive result $P_0$ as a function of the symbol intensity $\abs{\alpha}^2$. The analytic expression for the global bound is given in Eq.~(\ref{eqn:PT_single_degen}). (b) Communication capacity, $C$, as a function of the symbol intensity $\abs{\alpha}^2$. Linear receivers numerically saturate the globally optimal communication capacity computed using the theory of Peres and Terno~\cite{Peres1998_JPA}.}
\label{fig:single_degeneracy}
\end{figure*}%

\noindent
Random codes are typically non-degenerate. This means the coherent states in the random codes are almost surely linearly independent, i.e., the associated $R$ is full rank. We now consider examples of degenerate codes. As discussed above, our receiver, endowed with phase-space displacement, can map a code with single degeneracy into a non-degenerate one. As an example, consider the code with single-degeneracy~\cite{Guha2011} 
\begin{align}
\ket{\bm{\alpha}^1} = \ket{-\alpha,\alpha} \, , \,
\ket{\bm{\alpha}^2} = \ket{\alpha,-\alpha} \, , \, 
\ket{\bm{\alpha}^3} = \ket{\alpha,\alpha} \, .
\label{eq:Guha2011}
\end{align}
In Fig.~\ref{fig:single_degeneracy}(a), we illustrate the optimized inconclusive probabilities for linear receivers with displacement. Since these amplitude vectors are not linearly independent and have single degeneracy, receivers limited to LOP unitaries alone perform poorly, and we must leverage displacement operations. By allowing for phase-space displacement, our receivers match the global bound for small intensities ($e^{|\alpha|^2} < \sqrt{2}$). More details on the optimal receiver design are provided in Appendix \ref{app:rec}. The analytic expression of the global bound can be obtained using the method of Ref.~\cite{BergouPRL_2012}, which yields
\begin{align}
    P_0^\text{global} = \left\{
    \begin{array}{ccc}
    \frac{1}{3} \left( 2 e^{-4 |\alpha|^2} + 1 \right) & \mbox{if} & e^{|\alpha|^2} < \sqrt{2} \, , \\
    \frac{2}{3} \left( 2 e^{-2 |\alpha|^2} - e^{-4|\alpha|^2}   \right) & \mbox{if} & e^{|\alpha|^2} \geq \sqrt{2} \, .
    \end{array}
    \right.
\label{eqn:PT_single_degen}    
\end{align}%
The numerical optimality of linear receivers extends to larger signal intensities for alternative figures of merit, as discussed in the next Section.


\subsection{Alternative figures of merit}
\label{app:alternative_fom}

\noindent
The gap between linear receivers and the global bound can be further reduced for alternative figures of merit. Consider for example the communication capacity associated with the given input code words and a given detection strategy. The coherent states in Eq.~(\ref{eq:Guha2011}) can be used as code words for a communication protocol where the sender (Alice) uses these coherent states to encode a random variable $X$ that takes values $x=1,2,3$ with associated probabilities $p_X(x)$. The receiver (Bob) decodes this information by applying either the linear receiver or a globally optimal USD receiver. The outcome of the receiver is described by a random variable $Y$ that takes four possible values, $y=0,1,2,3$, with probability $p_Y(y)$, where $y=0$ is the inconclusive event. The maximum asymptotic communication rate achievable in this way is given by the Shannon capacity~\cite{Cover2006book}
\begin{align}
C & = \max_{p_X} H(Y) - H(Y|X) \, ,
\end{align}
where 
\begin{align}
H(Y) = - \sum_{y} p_Y(y) \log{p_Y(y)}
\end{align} 
is the Shannon entropy of $Y$, and 
\begin{align}
H(Y|X) = - \sum_x p_X(x) \sum_{y} p_Y(y|x) \log{p_Y(y|x)}
\end{align}
is the conditional entropy, where $p_Y(y|x)$ is the conditional probability of $Y=y$ given $X=x$.

Since the receivers are unambiguous, $P_0(x) := p_Y(y=0|x)$ is the probability of an inconclusive event for given input, and $p(y|x) = \delta_{yx}[1-P_0(x)]$ for $y=1,2,3$. The key quantity that determines the capacity is thus the conditional inconclusive probability $P_0(x)$. For linear receivers with displacement, this is given by 
\begin{align}
P_0(x) = \exp{\left[ - k_x \left| \bm{v}_x \cdot (\bm{\alpha}^x + \bm{\gamma}) \right| \right]} \, .
\end{align}
From this we obtain
\begin{align}
\begin{split}
C & = \max_{p_X} \left\{ - P_0 \log{P_0} - \sum_x p_X(x) \log{p_X(x)} \right. \\
&\hspace{1.7cm}+ \left. \sum_x p_X(x) P_0(x) \log{[ p_X(x) P_0(x) ]} 
\right\} \, ,
\label{rate}
\end{split}
\end{align}
where $P_0 = \sum_x p_X(x) P_0(x)$ is the average inconclusive probability. A comparison between our scheme (with LOP and displacement) and the global bound is shown in Fig.~\ref{fig:single_degeneracy}(b). The two schemes yield nearly equal capacities, the gap being too small to be visualized in the scale of the plot. 

\begin{figure}[t!]
\begin{center}
\includegraphics[width=0.98\columnwidth]{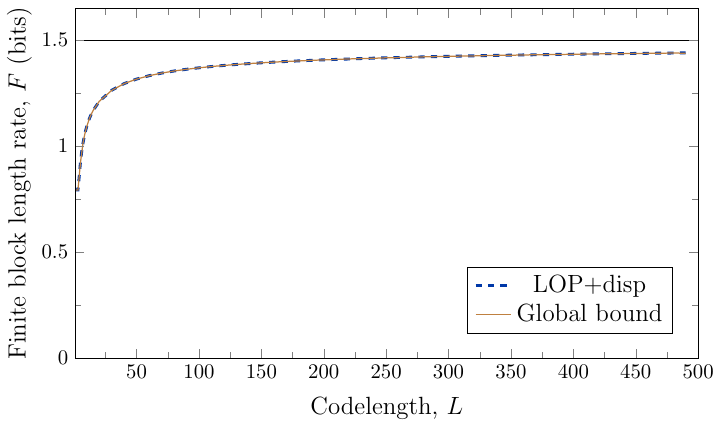}
\caption{Finite block length rate, $R$, in bits with varying code length $n$. Note that our scheme (dashed blue) with linear optics and single-mode detection saturates the global bound computed from the theory of Peres and Terno scheme (solid brown).}
\label{fig:fixed_CBs_finite_rate}
\end{center}
\end{figure}

In addition to the inconclusive probability and the Shannon capacity, we explore the optimality of our receiver using the finite communication block length rate, $F$. This rate is the communication rate attainable when both the code length, $L$, is finite, and there is a block error probability threshold, $\epsilon$, imposed on the communication~\cite{Polyanskiy2010_IEEE}. The normal approximation to the finite block length rate is given by~\cite{Tan2015_IEEE}
\begin{align}
F(L, \epsilon) = C - \sqrt{\frac{V}{L}} \, Q^{-1}(\epsilon),
\end{align}
where $\smash{Q(x) = 1/\sqrt{2\pi}\int^\infty_x dt \; \exp[-t^2/2]}$, and $V$ denotes the variance of the information transition probabilities of the channel,
\begin{align}
V = \sum_{x,y} p_Y(y) p_Y(y \vert x) \left(\log_2 \left[\frac{p_Y(y \vert x)}{\sum\limits_{z} p_Y(z) p_Y( z \vert x)}\right] - \overline{X} \right)^2 \, ,
\end{align}
and
\begin{align}
    \overline{X} = \sum_{x,y}  p_Y(y) p_Y(y \vert x) \log_2 \left[\frac{p_Y(y \vert x)}{\sum_{z} p_Y(z) p_Y(z\vert x)}\right] \, .
\end{align}
Figure~\ref{fig:fixed_CBs_finite_rate} illustrates the outcome of this maximization with different code lengths $L$ and for different receivers. We find that linear receivers match the finite-code-length performance generated from the global bounds. Note that in the asymptotic regime of large code block lengths, the finite rate tends towards the channel capacity with $\epsilon = 0$, which are illustrated in horizontal lines.

%


\subsection{Codes with double degeneracy}
\label{subsec:degenerate_codes-2}

\begin{figure*}[t!]
    \centering
    \includegraphics[width=\linewidth]{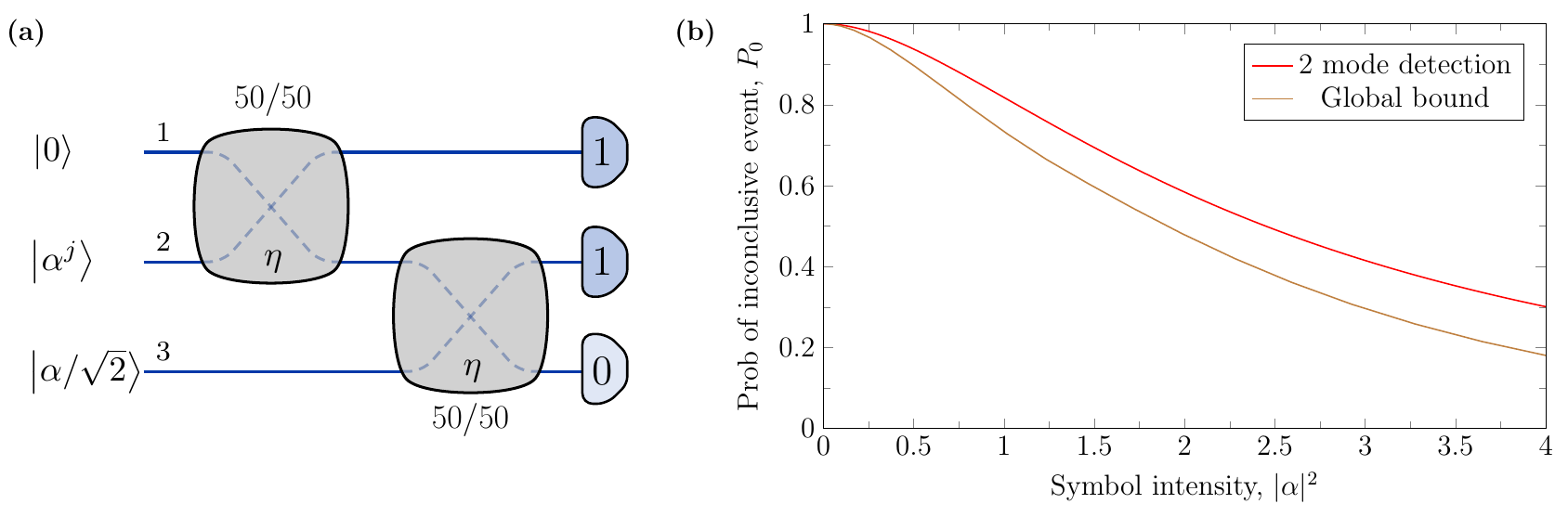}
    \caption{USD for the single-mode code with double degeneracy in Eq.~(\ref{eq:sing_mode_dd}): (a) Linear receiver implementing $50/50$ beam splitters ($\eta=1/2$) and joint on-off photodetection measurements, where mode $2$ carries the input coherent state, modes $1$ and $3$ carry auxiliary coherent states with amplitudes $0$ and $\alpha/\sqrt{2}$ respectively. For $\alpha^j=-\alpha$, clicks are only registered at the on-off photodetectors in modes 1 and 2, illustrated in darker blue. (b) The inconclusive event probability $P_0$ plotted versus $\abs{\alpha}^2$ for different receivers. The solid red line corresponds to our joint-detection receiver, and the solid brown line to the global bound for this code. The analytic expressions for the linear and the global bounds are given in Eq.~\eqref{eqn:dddecoder_performance} and Eq.~(\ref{eqn:PT_double_degen}) respectively.}
    \label{fig:double_degenerate_codes}
\end{figure*}

\noindent
To discriminate codes with higher degeneracy we may exploit joint detection events on pair of output modes. Consider the single-mode code with double degeneracy
\begin{align}
\ket{\alpha^1} = \ket{-\alpha} \, , \, 
\ket{\alpha^2} = \ket{\alpha} \, , \,
\ket{\alpha^3} = \ket{0} \, .
\label{eq:sing_mode_dd}
\end{align}
Single-mode codes find important applications in quantum sensing and communications protocols where individual measurements are preferred at each instance. Given the double degeneracy, our displacement-based receiver is not able to unambiguously discriminate signal states in this code. However, we can still discriminate these coherent states with LOP unitaries, vacuum auxiliary modes, displacement, and on-off detection by constructing receivers that exploit double detection events as described in Section~\ref{sec:2events}. The minimal inconclusive event probability of this class of receivers is determined as the solution to optimization ~(\ref{opt2}). By noting the equivalence between arbitrary auxiliary modes and vacuum auxiliary modes followed by mode-wise displacement operations, an explicit receiver that saturates the performance of linear receivers for this code is illustrated in Fig.~\ref{fig:double_degenerate_codes}(a). It requires two auxiliary modes: one prepared in the vacuum and the other in a coherent state with amplitude $\alpha/\sqrt{2}$ (as noted above using coherent state ancillas is equivalent to using vacuum ancillas and phase-space displacements). 
The modes are then mixed at two $50/50$ beam splitters and the output modes are measured by on-off photodetection. For an input amplitude $\alpha^j$ chosen from~\eqref{eq:sing_mode_dd}, we write the output coherent state amplitude at mode $k$ as $\zeta^j_k$. This input-output transformation on the coherent state amplitudes is summarised in Table~\ref{tab:signal_table}. By inspection of the table, it is evident that the coherent states can be unambiguously discriminated if two detectors click. For example, a joint detection on modes one and two identifies the input code word $\ket{\smash{{\alpha}^1}}=\ket{-\alpha}$ without error. From table~\ref{tab:signal_table}, the probability of a joint photodetection event in modes $k$ and $l$ can be determined through
\setlength{\thickarrayrulewidth}{2.1\arrayrulewidth}
\renewcommand{\arraystretch}{1.5}
\setlength{\tabcolsep}{8pt}
\begin{table}[t!]
  \centering
  \begin{tabular}{>{\centering\arraybackslash}m{1.9cm}||>{\centering\arraybackslash}m{1.38cm}|>{\centering\arraybackslash}m{1.32cm}|>{\centering\arraybackslash}m{1.32cm}}
    \thickhline
    \textbf{Input state} & \textbf{Output mode 1} & \textbf{Output mode 2} & \textbf{Output mode 3} \\
    \hline
        $|\alpha^1\rangle =  |-\alpha\rangle$ 	& $|-\alpha / \sqrt{2}\rangle$ 	& $\ket{\alpha}$ 		& $|0\rangle$             \\
        $|\alpha^2\rangle = |\alpha\rangle$  	& $|\alpha / \sqrt{2}\rangle$ 		& $|0\rangle$ 			& $ |-\alpha\rangle $    \\
        $|\alpha^3\rangle = |0\rangle$       	& $|0\rangle$ 					& $|\alpha/2\rangle$ 	& $|-\alpha/2\rangle$ \\
    \thickhline
  \end{tabular}
  \caption{Input-output transformation of the coherent state amplitudes through our linear receiver with two-mode photon detection. The receiver is shown in Fig.~\ref{fig:double_degenerate_codes}(a).}
  \label{tab:signal_table}
\end{table}%
\begin{align}
P(k, l\vert j) = (1-\exp[-\abs{\zeta^j_k}^2])(1-\exp[-\abs{\zeta^j_l}^2]) \, .
\end{align}
For this receiver, an inconclusive event corresponds to the absence of a joint photodetection event. 
Given a prior distribution $p^j$, the average probability of obtaining an inconclusive event is then given by
\begin{align}
\begin{split}
P_0 = 1 - \sum_{j=1}^3 p^j 
\sum_{l=1}^3 
\sum_{k < l}
P(k,l\vert j) \, ,
\end{split}
\end{align}
which for a uniform prior, $p^j=1/3$, yields
\begin{align}
\begin{split}
P_0
= \frac{1}{3} e^{-\frac{3 \abs{\alpha}^2}{2}} \left(e^{\abs{\alpha}^2} + 2 e^{\frac{\abs{\alpha}^2}{2}}+2 e^{\frac{5 \abs{\alpha}^2}{4}}-2\right),
\end{split}
\label{eqn:dddecoder_performance}
\end{align}
This result can be compared with the global bound, which we compute analytically using Ref.~\cite{BergouPRL_2012} (see Appendix~\ref{app:analytic_solution} for proof):
\begin{align}
    P_0^\text{global} = \left\{
    \begin{array}{ccc}
    \frac{1}{3} \left( 4 e^{-|\alpha|^2} - 2 e^{-2|\alpha|^2}  + 1 \right) & \mbox{if} & e^{|\alpha|^2} < 4 \, , \\
    \frac{2}{3} \left( 2 e^{-|\alpha|^2/2} - e^{-2|\alpha|^2}   \right) & \mbox{if} & e^{|\alpha|^2} \geq 4 \, .
    \end{array}
    \right.
\label{eqn:PT_double_degen}
\end{align}
The performance of the linear receivers is benchmarked with the global bound in Fig.~\ref{fig:double_degenerate_codes}(b). We note that the behavior of $P_0$ for small $|\alpha|^2$ is compatible with the quadratic law in Eq.~(\ref{P0-double}).

Finally, we note that while photon number resolving detectors (PNRDs) could help improve the performance of our detectors, we do not expect their use to replace the reliance on detection events across multiple modes. In the case of the example in Eq.~\eqref{eq:sing_mode_dd}, $\smash{\ket{\alpha^1}}$ and $\smash{\ket{\alpha^2}}$ differ only in phase and have the same photon statistics. Therefore PNRDs, by themselves, cannot be sufficient to discriminate these two states.


\subsection{Phase-shift keying}
\label{subsec:phasemod}

\noindent
A common way to encode information into coherent states is by phase modulation, i.e., phase-shift keying. Previous works have considered practical schemes for USD of phase-shifted coherent states, both with and without feedback~\cite{VanEnk2002_PRA, Becerra2013_NC, Izumi2020_PRL}. In this Section we analyze USD of phase-shifted coherent states using our linear receivers, re-obtaining some of the results of van Enk~\cite{VanEnk2002_PRA}. 

Binary phase-shift keying (BPSK) is a code with two states on one mode ($m=1$, $c=2$): 
\begin{align}
    |\alpha^1\rangle = |\alpha\rangle \, , \, \,  
    |\alpha^2\rangle = |-\alpha\rangle \, .
\end{align}
It is well-known how to optimally discriminate BPSK, here we express this method in the framework of our linear receivers. Indeed, to discriminate these states we can apply the strategy described in Section \ref{sec:diplace}. Note that these states are not linearly independent in phase space. To make them independent, first, we add an auxiliary vacuum mode, generating the two-mode states $|\alpha , 0\rangle$, $|-\alpha , 0\rangle$. Second, we displace the auxiliary mode by $\alpha$, which yields two coherent states with linear independent amplitude vectors, namely $|\alpha, \alpha\rangle$ and $|-\alpha, \alpha\rangle$. Finally, we note that the Gram matrix of these latter states is diagonal, thus the code is equivalent to the PPM code. Applying the argument of Section~\ref{subsec:PPM}, we conclude that the BPSK code is optimally discriminated with linear receivers.

\begin{figure}[t!]
\begin{center}
\includegraphics[width=0.98\columnwidth]{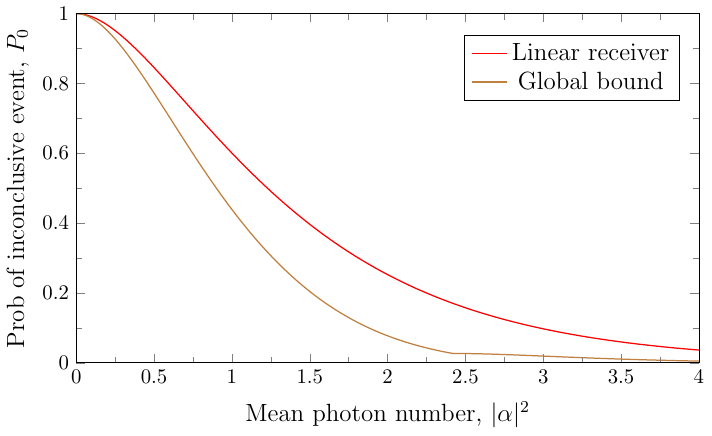}
\caption{The optimized inconclusive event probability $P_0$ plotted versus the 3PSK ensemble intensity $\abs{\alpha}^2$ for different receivers. The solid red line corresponds to the linear receiver, and the brown line to the global bound for this code.}
\label{fig:3PSK_doubledetection}
\end{center}
\end{figure}

In general, we may consider $M$-PSK codes with $M$ states over a single mode:
\begin{align}
    |\alpha^j\rangle = |\alpha \, e^{i j 2\pi/M}\rangle \, ,
\end{align}
for $j=0,\dots, M-1$. Note that for $M \geq 3$ the code has multiple degeneracies. This means that it can be discriminated only using multiple detection events. For example, the 3PSK ($M=3$) ensemble of coherent states can be discriminated by exploiting double detection events. For this, we make use of two auxiliary vacuum modes and construct linear receivers according to the strategy described in Section~\ref{sec:2events}. The best linear receiver saturates optimization~\eqref{opt2} and its performance is shown in Fig.~\ref{fig:3PSK_doubledetection} along with the global bound. Note that we recover the quadratic law of $P_0$ for small $|\alpha|^2$ as predicted by Eq.~(\ref{P0-double}).
For USD of 3PSK, it is easy to find the analytical expression for the optimized probability of the inconclusive event for linear receivers
\begin{align}
    P_0 = 1 
    - \left( 1 - e^{-|\alpha|^2 } \right)^2 \, .
\end{align}
Our numerical search suggests that using more than two vacuum auxiliary modes delivers no additional advantage to USD. Finally, the global bound, obtained from \cite{BergouPRL_2012}, reads
\begin{align}
 P_0^\text{global} 
 & = e^{-3|\alpha|^2/2}
 \max\left\{ 
 - 2 \cos{(\sqrt{3}|\alpha|^2/2)}
 , \right. \nonumber\\ 
 & \phantom{=====} \left. \cos{(\sqrt{3}|\alpha|^2/2)} \pm \sqrt{3} \, \sin{(\sqrt{3}|\alpha|^2/2)}
 \right\}  \, .
\end{align}

%


\section{Conclusions}
\label{sec:conc}

\noindent
While quantum states cannot in general be discriminated against without error, they can be discriminated against unambiguously if we allow for an inconclusive outcome. The theory of USD allows us to identify the globally optimal measurements and to compute the ultimate bounds on the probability of obtaining an inconclusive outcome, see for example~\cite{Peres1998_JPA,Bergou2001_PRA,BergouPRL_2012}. However, existing literature has rarely explored USD under realistic experimental constraints and limitations in what measurements are currently feasible, especially for the discrimination of coherent states of the quantum electromagnetic field.

Here we have outlined a theory of USD for multimode coherent states that focuses on practical receivers that can be feasibly realized with current technologies. Our receivers operate under physical resources described entirely through linear optical passive (LOP) unitaries, phase-space displacement operations, auxiliary vacuum modes, and on-off mode-wise photon detection. We have benchmarked the performance of these linear receivers against the global bound and found examples where they are optimal or near-optimal. In particular, this happens for random codes. Our findings show that high-performance USD receivers can be readily realized with currently available technologies, and suggests that, at least for typical, non-degenerate codes, high-order non-linearities, feedback, or more advanced quantum technologies may only provide small improvements.

We do not have a complete theoretical explanation of why linear receivers perform well. However, we think that the reason may be related to the fact that, at least for small amplitudes, the probability of multiple detection events is suppressed. Therefore, using a photon number resolving detector may not be particularly useful unless the code is degenerate. Furthermore, as the photon number is subject to quantum fluctuations in coherent states, it is unlikely that more detailed information on photon statistics can improve USD (whereas it may be useful for ASD). Also, we have seen that linear receivers achieve the global bound for the PPM codes, and for all codes that can be reduced to them by linear optics. By symmetry, and by an argument of concentration of measure, we may indeed expect that, up to statistical fluctuations, a random code is not too different from a PPM code: this is at least true in the regime where the number of modes ($m$) is much larger than the number of code words ($c$). In fact in this regime the scalar product $(\bm{\alpha}^i,\bm{\alpha}^j)$ (which is zero on average) has fluctuations of order $1/\sqrt{m}$ and the Gram matrix becomes nearly diagonal if $c \ll m$. However, for small values of $m$, the displacement operation seems to play an important role that is not captured by this argument.

A number of questions remain open. First, an analytical expression for the inconclusive probability, beyond our numerical results, would characterize more clearly the comparison with the global bound, and allow us to apply our analysis to an arbitrary number of modes and code words. In the regime of weak signals, this problem may be more naturally framed within the language of Poisson quantum information~\cite{Poisson}. Second, an extension of our theory is required to handle highly degenerate codes, when feedback operations could be useful. This is especially important when the number of states is much larger than the number of optical modes. In this regime, which is of particular interest for quantum communications, USD may be achieved only by exploiting joint photodetection events on multiple modes. Finally, our work might be formulated as a resource theory, similar to Refs.~\cite{Winter2021,Lami2021}.
This approach may provide insight and help in comparing different sets of resources, for example by including homodyne or heterodyne detection, photon addition, and subtraction, or some mild non-linear interactions.

\vspace{0.2cm}

\section{Acknowledgments}
\noindent
This work was supported by the EPSRC Quantum Communications Hub, Grant No.~EP/T001011/1,
and from the European Union's Horizon Europe research and innovation programme under the project ``Quantum Secure Networks Partnership'' (QSNP, grant agreement No.~101114043. 
SG acknowledges the NSF Center for Quantum Networks, awarded under grant number EEC-1941583. 
MSB acknowledges the University of Arizona's {\em Information in a Photon} course for being introduced to the problem of optimal linear-optic USD receiver designs. 
CL acknowledges financial support from PNRR MUR project PE0000023-NQSTI. We are grateful to the anonymous referees for their insightful and stimulating comments.


\appendix

\section{Peres-Terno theory for USD}
\label{sec:USD_theory}

\noindent
In this Appendix we review the theory of Peres and Terno~\cite{Peres1998_JPA} for the USD of a set of linearly independent vectors $\ket{\smash{u_j}}$ with $j=\{1,\dots,c\}$, associated with prior probabilities $p^j$ (satisfying $\sum_{j=1}^c p^j =1$). These vectors span a $c$-dimensional Hilbert space $\mathcal{H}_c$. We define a unique set of (not necessarily normalized) vectors $\ket{\smash{v_j}} \in \mathcal{H}_c$ such that
\begin{align}
    \braket{ v_i \vert u_j} = \delta_{ij} \braket{ v_i \vert u_j}  \, .
\label{vec_overlap}    
\end{align}
We use these vectors to define a POVM with $n$ elements
\begin{align}
    A_j = k_j^2 \ket{\smash{v_j}} \bra{\smash{v_j}} \, ,
\label{eqn:POVMj}    
\end{align}
for $j=1,\dots,n$. The POVM element corresponding to an inconclusive event is
\begin{align}
    A_0 = I - \sum_{j=1}^c k_j^2 \ket{v_j} \bra{v_j } \, .
\label{eqn:POVM0}    
\end{align}
The parameters $k_j$ are chosen in such way to ensure $A_0 \geq 0$, and $I$ is the identity in $\mathcal{H}_c$. For a suitable choice of the parameters $k_j$'s, this POVM allows for unambiguous discrimination. The corresponding probability of an inconclusive outcome is 
\begin{align}
    P_0 = \sum_{j=1}^c p^j \braket{ u_j \vert A_0 \vert u_j }
        = 1 - \sum_{j=1}^c p^j k_j^2 \abs{\braket{ u_j \vert v_j }}^2 \, .
\end{align}
A globally optimal USD measurement corresponds to the one that minimizes $P_0$ subject to the positivity of $A_0$. This is obtained as the solution to the constrained maximization problem:
\begin{maxi}
{k_1,\ldots,k_n}{\sum_{j=1}^c p^j k_j^2 |\langle u_j | v_j \rangle|^2 \, ,}
{}{}
\addConstraint{\sum_{j=1}^c k_j^2 |v_j \rangle \langle v_j |}{\leq I \, ,}
\label{opt0_PT}
\end{maxi}
which defines the globally optimal bound on the probability of the inconclusive outcome.


\section{Method of Bergou, Futschik, and Feldman}
\label{sec:Bergou}

\noindent
Consider the $c \times c$ matrix with elements
\begin{align}
    C_{ij} = \langle u_i | A_0 | u_j \rangle \, .
\end{align}
Note that the diagonal elements are the probabilities of an inconclusive event conditioned on the vector $|u_j\rangle$,
\begin{align}
    q_j = C_{jj}\, ,
\end{align}
and the off-diagonal elements are simply the overlaps between the code words, 
\begin{align}\label{offd}
    C_{ij} = \langle u_i | u_j \rangle \, ,
\end{align}
for $i \neq j$.

If the vectors $|u_j\rangle$ are linearly independent, the condition of non-negativity of the operator $A_0$, $A_0 \geq 0$, is equivalent to $C \geq 0$. Therefore, following the theory of Bergou et al.~\cite{BergouPRL_2012}, the optimal average inconclusive probability is obtained by solving the constrained optimization
\begin{mini}
{q_1,\ldots,q_j}{\sum_{j=1}^c p^j q_j \, ,}
{}{}
\addConstraint{C }{\geq 0 \, .}
\label{opt_Bergou}
\end{mini}
Note, the solution to this optimization when $q_j = q$ for all $j$ is known to be the minimal eigenvalue of the Gram matrix, that is, $q$~\cite{Sentis2017_PRL}. More generally, for non-equal $q_j$, this problem can be solved analytically or semi-analytically. The case of $c=3$ is discussed in detail in Ref.~\cite{BergouPRL_2012}.

We can immediately apply this method to show that PPM codes are optimally discriminated by mode-wise photodetection. To prove this, recall that the PPM codes are defined such that the matrix $R$ is square, with $R=\alpha I$ and $I$ the $c$-dimensional identity matrix. The off-diagonal entries of the matrix $C$ are all equal to $\smash{e^{-|\alpha|^2}}$, while the diagonal entries are all equal, i.e., $C_{jj} = q$. Therefore, the objective function in the minimization~\eqref{opt_Bergou} is equal to $q$. The eigenvalues of $C$ are $\smash{(c-1) e^{-|\alpha|^2} + q}$ (with multiplicity $1$) and $\smash{q-e^{-|\alpha|^2}}$ (with multiplicity $c-1$). The smallest value of $q$ such that the eigenvalues remain non-negative is therefore $\smash{q = e^{-|\alpha|^2}}$, which matches the inconclusive event probability obtained through mode-wise photodetection in Section~\ref{subsec:PPM}.


\section{Framework for optimized USD receivers}
\label{sec:Mconstraint}

\noindent
In this Appendix, we determine a condition for a $c \times m$ matrix $M$ (with $c\leq m$) to be a submatrix of a larger unitary matrix $U$. First, if $c<m$, extend $M$ into a square, $m \times m$ matrix, $M_0$ by appending $m-c$ rows of zeros. Second, apply the singular value decomposition $M_0 = \mathcal{U} D \mathcal{V}$, where $\mathcal{U}$ and $\mathcal{V}$ are unitary matrices, and $D$ is diagonal with non-negative entries. For $D \leq 1$, the following $2 m \times 2 m$ matrix is unitary:
\begin{align}
V =    \left( \begin{array}{c|c}
    D   & -\sqrt{I-D^2} \\
    \hline
    \sqrt{I-D^2} & D 
    \end{array}
    \right) \, ,
\end{align}
where $I$ is the identity matrix. By multiplying $V$ by $\mathcal{U}$ and $\mathcal{V}$ we obtain another unitary matrix, which is a unitary extension of $M_0$,
\begin{align}
\begin{split}
    U &= \left( \begin{array}{c|c}
    \mathcal{U} & 0 \\
    \hline
    0           & I 
    \end{array}
    \right)
    \left( \begin{array}{c|c}
    D   & -\sqrt{I-D^2} \\
    \hline
    \sqrt{I-D^2} & D 
    \end{array}
    \right)
        \left( \begin{array}{c|c}
    \mathcal{V} & 0 \\
    \hline
    0           & I 
    \end{array}
    \right) \\
    &=
    \left( \begin{array}{c|c}
    M_0   & - \mathcal{U} \sqrt{I-D^2} \\
    \hline
    \sqrt{I-D^2} \, \mathcal{V} & D 
    \end{array}
    \right) \, .
\end{split}
\end{align}
As $M_0$ is an extension of $M$, it follows that $U$ is a unitary extension of $M$. We conclude that a matrix $M$ can be extended into a unitary, if and only if its singular eigenvalues are not larger than $1$. This condition can equivalently be written as
\begin{align}\label{M1}
M^\dag M \leq I \, , \, \, \mbox{or} \, \, M M^\dag \leq I \, .
\end{align}
Note that with this construction, the unitary matrix has at most size $2m$. For our application, this means that we need at most $m$ auxiliary modes to implement the receiver. In particular, the minimum number of auxiliary modes equals the number of singular values of $M_0$ that are strictly smaller than $1$.


\section{Receiver for a code with single degeneracy}
\label{app:rec}

\begin{figure}[h!]
\centering
\includegraphics[width=0.55\linewidth]{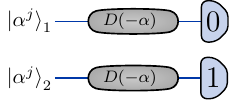}
\caption{Optimal receiver (in low-photon regime with $\abs{\alpha}^2 \leq \ln \sqrt{2}$) for the two-mode, single-degeneracy CB~\eqref{eq:Guha2011_2}, comprised of displacing each mode by $-\alpha$ before on-off photodetection.}
\label{fig:s0}
\end{figure}

\noindent
Here we discuss in more detail the optimal linear-optics receiver for the code in Eq.~(\ref{eq:Guha2011})
\begin{align}
\ket{\smash{\bm{\alpha}^1}} = \ket{-\alpha,\alpha} \, , \, 
\ket{\smash{\bm{\alpha}^2}} = \ket{\alpha,-\alpha} \, , \, 
\ket{\smash{\bm{\alpha}^3}} = \ket{\alpha,\alpha} \, .  
\label{eq:Guha2011_2}
\end{align}
By applying a displacement $\bm{\gamma} = (-\alpha,-\alpha)$ we can map the third coherent state into the vacuum, and the first two into a PPM code:
\begin{align}
\ket{\smash{\bm{\beta}^1}} = \ket{-2\alpha,0} \, , \, 
\ket{\smash{\bm{\beta}^2}} = \ket{0,-2\alpha} \, , \, 
\ket{\smash{\bm{\beta}^3}} = \ket{0} \, .  
\end{align}
Each time a photo-detection event is observed, we have un-ambiguous discrimination. This scheme is shown in Fig.~\ref{fig:s0}. Note that in this way, the third coherent state is never discriminated. It is easy to check that the probability of the inconclusive event achieved in this way is $P_0 = \frac{1}{3} \left( 2 e^{-4|\alpha|^2} + 1 \right)$ and thus, comparing with Eq.~(\ref{eqn:PT_single_degen}), the scheme is optimal for $e^{|\alpha|^2} < \sqrt{2}$.

\begin{figure}[h!]
\centering
\includegraphics[width=0.9\linewidth]{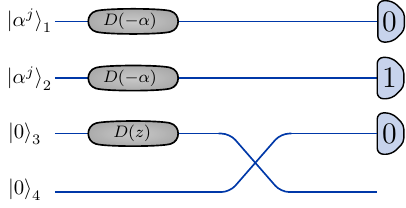}
\caption{Optimal receiver (in low-photon regime with $\abs{\alpha}^2 \leq \ln \sqrt{2}$) for the two-mode, single-degeneracy CB~\eqref{eq:Guha2011_2}, comprised of two vacuum auxiliary modes, displacement across the first three modes, a swap operation between modes three and four, before on-off photodetection.}
\label{fig:s1}
\end{figure}

To frame this scheme into our theory, we need to add an ancillary mode, initially in the vacuum state, then displace the three modes by $\bm{\gamma} = (-\alpha, -\alpha, z)$. We then add another vacuum ancillary mode and swap the third and fourth modes. USD is then achieved by the photodetection of the first three modes. This is shown in Fig.~\ref{fig:s1}. Note that the third detector never clicks, which is expected as the third code word is not discriminated in this scheme.

\begin{figure}[h!]
\centering
\includegraphics[width=\linewidth]{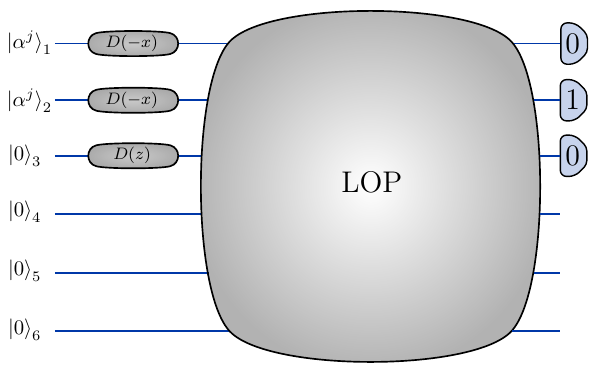}
\caption{Near-optimal receiver (in the higher-photon regime with $\abs{\alpha}^2 > \ln \sqrt{2}$) for the two-mode, single-degeneracy CB~\eqref{eq:Guha2011_2}.}
\label{fig:s2}
\end{figure}

For larger values of $\alpha$, i.e.,~for $e^{|\alpha|^2} > \sqrt{2}$, LOP receivers are only near-optimal. Our numerical search indicates that the optimal displacement is of the form $\bm{\gamma} = (-x,-x,z)$, including one ancillary vacuum mode, where the positive parameter $x$ is in general smaller than $|\alpha|$.
It follows that the unit vectors $\bm{v}_i$ in (\ref{opt1}) are
\begin{align}
\bm{v}_1 & = \mathcal{M} ( z , 0 , x-\alpha ) \, , \\
\bm{v}_2 & = \mathcal{M} ( 0 , z , x-\alpha ) \, , \\
\bm{v}_3 & = \mathcal{N} ( z, z , 2 x ) \, ,
\end{align}
where $\mathcal{M}$ and $\mathcal{N}$ are normalization factors. Numerically, we find we can assume without loss of generality $k_1 = k_2 = k_3 =: k$. Therefore, the matrix $M$ reads 
\begin{align}
M =
\left( 
\begin{array}{ccc}
 k \mathcal{M} z & 0 & k\mathcal{M}(x-\alpha)  \\
  0 &  k \mathcal{M} z & k\mathcal{M}(x-\alpha) \\
 k \mathcal{N} z & k\mathcal{N} z &  2 k \mathcal{N} x 
\end{array}
\right) \, ,
\end{align}
which in turn can be completed into a $6 \times 6$ (real-valued) unitary matrix following for example the procedure of Appendix \ref{sec:Mconstraint}. This scheme is shown in Fig.~\ref{fig:s2}.

%


\section{Example of code with double degeneracy}
\label{app:analytic_solution}

\noindent
In this Appendix, we analytically derive the global bound for the following family of single-mode coherent states with double degeneracy
\begin{align}
\ket{\smash{\alpha^1}} = \ket{-\alpha} \, , \, 
\ket{\smash{\alpha^2}} = \ket{\alpha} \, , \,
\ket{\smash{\alpha^3}} = \ket{0} 
\end{align}
using the method and notation of Bergou~\emph{et al.}~\cite{BergouPRL_2012}. Writing the overlaps $\smash{\braket{\alpha^1\vert \alpha^2}=s_3e^{i\phi_3}}$ with two other cyclic permutations of the indices for $0 \leq s_j \leq 1$, we have
\begin{align}
    s_{3} & = \braket{ \alpha^1 |\alpha^2} = e^{-2|\alpha|^2} \, , \\
    s_{1} & =\braket{ \alpha^2 |\alpha^3 } = e^{-|\alpha|^2/2}
    = s_2 \, ,
\end{align}
with $s_j \in \mathbbm{R}$ for $j\in\{1,2,3\}$, and $\phi_j=0$. Hence, the global invariant phase $\phi=\sum_j\phi_j=0$. Writing $s := s_1 = s_2$, Bergou \emph{et al.} introduce the cyclic parameters
\begin{align}
    r_3 & = \frac{s_3}{s_1 s_2} = e^{-|\alpha|^2} , \; r  := r_1 = r_2 = \frac{1}{s_3} = e^{2|\alpha|^2} \, .
\end{align}
The minimized total failure probability from Eq.~\eqref{opt_Bergou} can then be written as an optimization over the scaled failure probabilities $\smash{\tilde q_j = r_j q_j \leq r_j}$~\cite{BergouPRL_2012}:
\begin{mini} 
{\tilde q_1, \tilde q_1, \tilde q_3}{Q = \sum_{j=1}^3 \eta_j \frac{\tilde q_j}{r_j} \, ,}
{}{}
\addConstraint{0 \leq \tilde q_j}{\leq r_j} 
\addConstraint{\tilde q_j \tilde q_k - 1}{\geq 0}
\addConstraint{\tilde q_1 \tilde q_2 \tilde q_3 - \tilde q_1 - \tilde q_2 - \tilde q_3 + 2\cos\phi}{\geq 0}\, .
\label{eqn:bergou_results}
\end{mini}
Note that the first constraint follows from the positivity of $\smash{q_j}$, the second from the positivity of the failure POVM $\Pi_0$, and the third from the linear independence (LI) of the states. Collectively, these constraints ensure the non-negativity of $C$ in Eq.~\eqref{opt_Bergou}.

\begin{figure}[b!]
\centering
\includegraphics[width=0.98\linewidth]{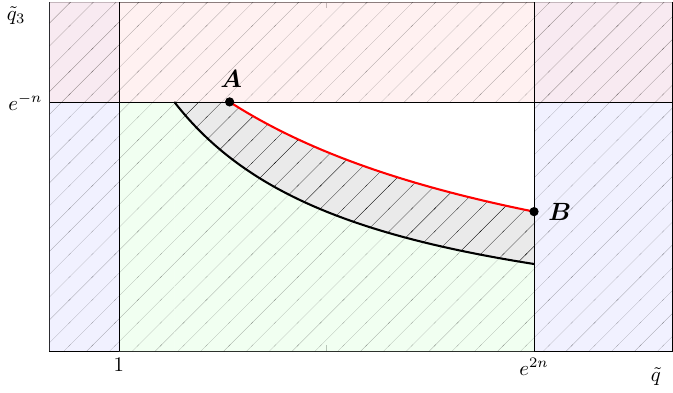}
\caption{Feasible domain in the $\tilde q-\tilde q_3$ space shown in white:
The lower (black) curve is $\smash{\tilde q_3 = \tilde q^{-1}}$.
The upper (red) curve is $\smash{\tilde q_3 = 2(\tilde q + 1)^{-1}}$. The coordinates $(\tilde{q}, \tilde{q}_3)$ of vertices $A$ and $B$ are $\smash{(2 e^{|\alpha|^2} - 1, e^{-|\alpha|^2})}$ and $\smash{(e^{2|\alpha|^2}, 2(e^{2 |\alpha|^2}+1)^{-1})}$ respectively.
The unshaded (white) region, referred to as the interior region, corresponds to the optimal parameter space for $\{\tilde{q}, \tilde{q}_3\}$.}
\label{fig2}
\end{figure}

To analytically solve~\eqref{eqn:bergou_results}, note that the failure scaled probabilities satisfy $\smash{\tilde q := \tilde q_1 = \tilde q_2}$. This symmetry reduces the general three-dimensional feasible region in $\{\tilde q_1, \tilde q_2, \tilde q_3\}$ to a two-dimensional feasible region in $\{\tilde q, \tilde q_3\}$ defined by a cross-section along the plane $\smash{\tilde q_1=\tilde q_2}$. The optimization then amounts to finding the minimum solutions to the objective function along the interior, edge, and vertices that defines this two-dimensional feasible constraint region. To see this explicitly, note that the constraints in~\eqref{eqn:bergou_results} constrain the feasible range for $\tilde q$, $\tilde q_3$ that collectively satisfy
\begin{align}
    \tilde q  \in [1,e^{2|\alpha|^2}] \, , \;
    \tilde q_3  \in [ 0 , e^{-|\alpha|^2} ] \, , \;
    \tilde q_3  \geq \frac{2}{\tilde q + 1}  \, .
\end{align}
This feasible region for $\{\tilde q, \tilde q_3\}$ is illustrated within the 2D unshaded area in Fig.~\ref{fig2}. 

\begin{figure*}[t!]
\centering
\includegraphics[width=1\linewidth]{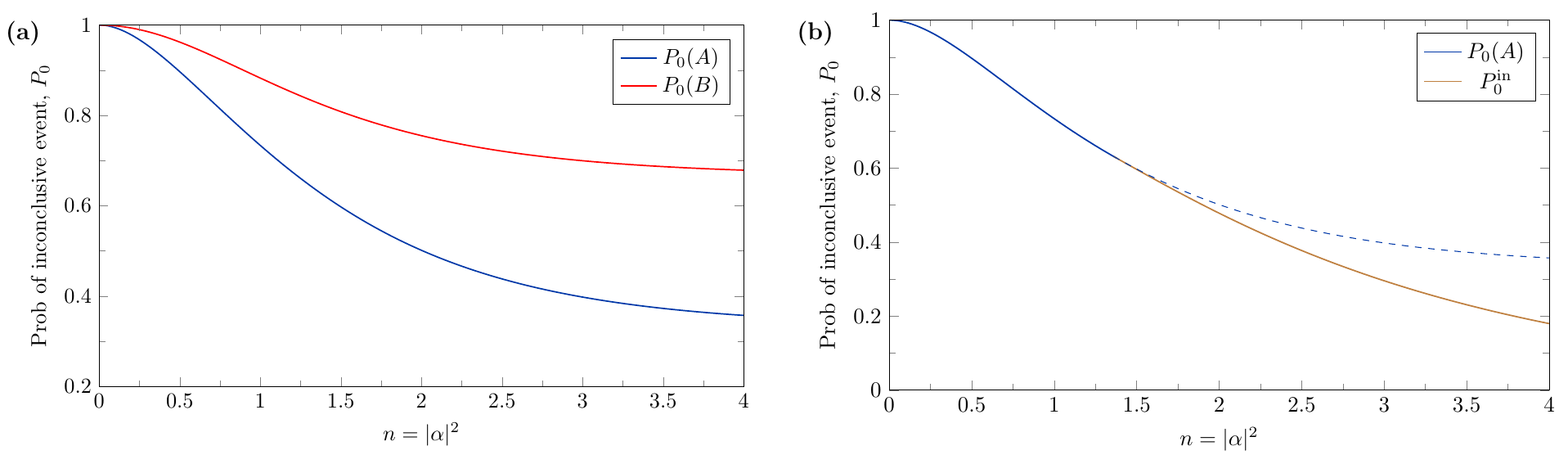}
\caption{Edge and vertex solutions for minimized $P_0$: a) The solution $P_0(A)$ in blue, and the solution $P_0(B)$ in red, plotted versus $|\alpha|^2$. This shows that only vertex $A$ needs to be considered. b) The global minimum $P_0^\text{Global}$ as a function of $n$. The interplay between minimal inconclusive event probabilities as given by $P_0(A)$ in blue and $P_0^\text{in}$ in brown means the minimized solution given by $P_0(A)$ for $n\geq \ln (4)$ (dashed blue) is superseded by the solution given by $P_0^\text{in}$ (solid brown).}
\label{fig:AB_final_sols}
\end{figure*}

The objective function in Eq.~\eqref{eqn:bergou_results} to minimize in the allowed domain is (uniform \emph{a priori} distribution)
\begin{align}
    P_0(\tilde q, \tilde q_3) & = \frac{1}{3} \sum_{j=1}^3 \frac{\tilde q_j}{r_j} 
     = \frac{1}{3} \left( 2 \, e^{-2|\alpha|^2} \, \tilde q  + e^{|\alpha|^2} \, \tilde q_3  \right) \, .
\end{align}
Small values of $\tilde q$ and $\tilde q_3$ make this quantity smaller, this means that the minimum is on the curve that joins the points $A$ and $B$, as shown in Fig.~\ref{fig2}. We first compute $P_0$ on vertex $A$, which has coordinates $\smash{(\tilde q,\tilde q_3) = (2 e^{|\alpha|^2} - 1, e^{-|\alpha|^2})}$, and vertex $B$, with coordinates $\smash{(e^{2|\alpha|^2}, 2(e^{2|\alpha|^2}+1})^{-1})$, to give
\begin{align}
    P_0(A) &= \frac{1}{3} \left( 4 e^{-|\alpha|^2} - 2 e^{-2|\alpha|^2}  + 1 \right)\, ,\\
    P_0(B) &= \frac{1}{3} \left( 2 +  \frac{2 e^{|\alpha|^2}}{e^{2|\alpha|^2}+1} \right)\, .
\end{align}
A comparison of the two values is illustrated in Fig.~\ref{fig:AB_final_sols}(a), which shows that $P_0(A) \leq P_0(B)$. Therefore, we discard the point $B$ from the search for the minimized $P_0$ value.

Next, we minimize $P_0$ over the interior points between $A$ to $B$ along $\smash{\tilde q_3 = 2(\tilde q + 1)^{-1}}$:
\begin{align}
    P_0(\tilde q) = \frac{2}{3} \left( e^{-2|\alpha|^2} \, \tilde q  +  \frac{e^{|\alpha|^2}}{\tilde q + 1}  \right) \, ,
\end{align}
which amounts to a single parameter optimization. This function has a stationary point at $\smash{\tilde q_\text{in} = e^{3|\alpha|^2/2}-1}$ yielding
\begin{align}
    P_0^\text{in} = \frac{2}{3} \left( 2 e^{-|\alpha|^2/2} - e^{-2|\alpha|^2}   \right) \, .
\end{align}
The feasibility of this solution is conditioned on the stationarity point $\smash{q_\text{in}}$ residing within the allowed domain between vertices $A$ and $B$, which holds if and only if
\begin{align}
    e^{3|\alpha|^2/2}-1 \in [ 2 e^{|\alpha|^2} - 1, e^{2|\alpha|^2} ]
    \implies
    e^{|\alpha|^2} \geq 4 \, .
\end{align}
This solution is illustrated together with $P_0(A)$ in Fig.~\ref{fig:AB_final_sols}(b). The solution $P_0^\text{in}$ is plotted only when $e^{|\alpha|^2} \geq 4$. The two solution crosses exactly at $|\alpha|^2 = \ln(4)$. Hence, the global minimum for the double-degeneracy (DD CB is characterized by $P_0(A)$ for $e^n < 4$, and $P_0^\text{in}$ for $e^n \geq 4$:
\begin{align}
    P_0^\text{Global-DD} = \left\{
    \begin{array}{ccc}
    \frac{1}{3} \left( 4 e^{-|\alpha|^2} - 2 e^{-2|\alpha|^2}  + 1 \right) & \mbox{if} & e^{|\alpha|^2} < 4 \\
    \frac{2}{3} \left( 2 e^{-|\alpha|^2/2} - e^{-2|\alpha|^2}   \right) & \mbox{if} & e^{|\alpha|^2} \geq 4 \, ,
    \end{array}
    \right.
\end{align}
which concludes our proof. 
%


%

\end{document}